\documentclass[journal=jctcce,manuscript=article,layout=traditional]{achemso}
\usepackage[T1]{fontenc}
\usepackage[version=3]{mhchem}
\usepackage[dvipsnames]{xcolor}
\usepackage{graphicx}
\usepackage{subcaption}
\usepackage{amssymb}
\usepackage{overpic}
\usepackage{braket}
\usepackage{soul}
\usepackage{bm}
\usepackage{enumerate}
\usepackage{hyperref}

\hypersetup{pdfborderstyle={/S/U/W 0.5}}
\newcommand{\grad}[1]{\nabla_{\bm{#1}}}

\newcommand{\br}{{\mathbf r}}
\newcommand{\Br}[0]{{\bm{B}^{\rm r}}}
\newcommand{\Bi}[0]{{\bm{B}^{\rm i}}}
\newcommand{\Dr}[0]{{\bm{D}^{\rm r}}}
\newcommand{\Di}[0]{{\bm{D}^{\rm i}}}

\title[]{Charge Transfer with a Spin. I: A Variational Constrained-CASSCF Framework for Investigating Charge Transfer in the Presence of Spin--Orbit Coupling}
\author{Alok Kumar}
\affiliation{Department of Chemistry, University of Pennsylvania, Philadelphia, Pennsylvania, 19104, USA}
\alsoaffiliation{Department of Chemistry, Princeton University, Princeton, New Jersey, 08540, USA}
\author{Zhen Tao}
\affiliation{Department of Chemistry, University of Rhode Island, Kingston, Rhode Island, 02881,
USA}
\author{Joseph E. Subotnik}
\affiliation{Department of Chemistry, Princeton University, Princeton, New Jersey, 08540, USA}
\author{Tian Qiu}
\email{tianq@princeton.edu}
\affiliation{Department of Chemistry, Princeton University, Princeton, New Jersey, 08540, USA}

\begin{document}

\maketitle
\begin{abstract}
    Charge transfer in open-shell molecular systems can involve a delicate interplay between
    charge localization, orbital relaxation, and spin--orbit coupling (SOC), particularly near
    ground-excited state crossings.
    Here, we introduce a very inexpensive variational framework for treating these effects in odd-electron systems by extending the electron/hole-transfer
    Dynamically-weighted State-Averaged Constrained CASSCF (eDSC/hDSC) method to include SOC.
    Our method incorporates the SOC Hamiltonian directly into the variational orbital optimization through complex-valued spinor orbitals, allowing orbital and spin degrees of freedom to relax self-consistently while preserving the time-reversal symmetry of doublet states.
    The method achieves smooth potential energy surfaces and rapid self-consistent field (SCF) convergence; moreover, from physically small to artificially large SOC strengths, the calculation can be converged to very tight thresholds across the whole potential energy surface.
    While the results presented here are for systems with two charge centers, as shown in the appendix, the theory is quite general and can be adapted to either a system of $N$ charge centers or a metal-molecule surface with a continuum of states, so that charge transport  (and not just charge transfer) can also be studied.
    The approach therefore provides a route towards \emph{ab initio} studies of spin-dependent charge transfer and charge transport in molecular systems with nontrivial spin degrees of freedom.
\end{abstract}

\section{Introduction}
Charge transfer (CT) processes are ubiquitous in nature, playing key roles in photosynthesis\cite{meech1986role, wasielewski1992photoinduced} and cellular energy conversion processes\cite{moller2001plant, massey1994activation, siedow1995plant}; CT processes are also at the heart of many modern device applications involving organic electronics\cite{forrest2007introduction, deibel2010polymer, coropceanu2007charge} and light emitting diodes\cite{zhu2009charge, ostroverkhova2016organic, etherington2016revealing, wong2017purely, zhang2023highly}.  
 Unluckily for the experimentalist (but luckily for the theorist), CT is a very complicated process, the modeling of which is a full time job. On the one hand, it is well known that, in order to maintain energy conservation during a charge transfer process, the electronic and nuclear degrees of freedom must exchange energy; in other words, nuclear dynamics and  rearrangement cannot be avoided\cite{marcus1956,Hush:1961,Levich:1961,Cukier:1998,Jortner:1998,Dennerl:2010, penfold2018spin}. On the other hand, by definition, charge transfer processes inevitably involve multiple electronic states and therefore demand an accurate description of static electron-electron correlation--a challenging task, especially for thermal electron transfer, where we must balance the energies of ground and excited states\cite{helmich2019benchmarks, Athavale2021}.  While a great deal of progress has been made in modeling {\em ab initio} CT processes, and surface hopping techniques are now routine as far as simulating the nonadiabatic dynamics (the first problem)\cite{subotnik2016understanding,crespo2018recent, tapavicza2007trajectory, mai2018nonadiabatic}, performing efficient electronic structure calculations capable of treating multiple charge sites with just enough accuracy remains difficult
and is almost always the computational bottleneck preventing large scale  simulation \cite{gonzalez2012progress, lischka2018multireference, matsika2021electronic, dergachev2023predicting}.  

A few words are now appropriate about how one usually picks out an electronic structure algorithm for running nonadiabatic dynamics to simulate charge transfer, especially for thermal electron transfer or transport, which is the focus of the present paper. If one seeks to go beyond semi-empirical calculations \cite{tretiak:2014:acr}, for large scale simulations, one would like to simply model excited states as orbital energies. This approach has been followed for years by Prezhdo and co-workers with some success in the solid state, but it is worth noting that for molecules, orbital energies experience trivial crossings routinely and virtual orbital energies can be very sensitive to basis set \cite{prezhdo2021modeling, mei2019charge, fischer2011regarding}.  For two state dynamics, simulating nonadiabatic processes on the ground DFT and excited state TD-DFT states is common nowadays\cite{tapavicza2007trajectory,herbert2022spin, bernard2012general}, but the downside of such an approach is that when the two surfaces come together, one fails to recover the correct topology of a conical intersection\cite{Shao:2003:spin_filp, martinez:2006:ci_topology_wrong, taylor2023description}.  Interestingly, to our knowledge, very few (if any) nonadiabatic simulations have been run with constrained DFT\cite{wu:2007:cdftci, wu:2006:cdft, vv:2006:jcp}, which would provide another approach for balancing ground and excited states if merged with configuration interaction (but  which might fail for large couplings\cite{vv:2015:jcp_cdft_fails}). One appealing option for modeling radical thermal charge transfer is to use the cation or anion closed-shell molecule as a reference and then add or take away an electron (just like one might use a triplet reference for a spin-flip calculation\cite{casanova2020spin}). The resulting EOM-IP/EOM-EA simulations have been developed  by Krylov \emph{et al.} \cite{krylov2008equation,
pokhilko2019general}, and SOC as well as nuclear gradients are now available in Q-Chem electronic structure package \cite{epifanovsky2015spin,epifanovsky2021software,zhao2026analytic}. That being said, just as for spin-flip calculations, one must expect that the accuracy of the results will depend on how close is the relevant cation/anion electronic structure to the electronic structure of the radical, and large  diagonalizations may be needed.  
For the most part, as far as simulating nonadiabatic dynamics is concerned today, the most common approach nowadays is to run complete active space self-consistent field (CASSCF) simulations, where one optimizes both the orbital basis and the  wavefunction coefficients in that basis at the same time \cite{roos:1980, siegbahn1981complete}.  When properly state-averaged, CASSCF can correctly balance ground and excited states, and thus CASSCF would appear to be the most natural electronic structure approach for running nonadiabatic simulations.

In this paper, our interest is in exploring charge transfer processes between open-shell (radical) species where spin is involved.  For such a problem, CASSCF has three points of interest that must now be addressed.

\subsection{Point of Interest \#1: Globally Smooth Surfaces}

Point  of interest \#1 is that  CASSCF very often produces discontinuous potential energy surfaces\cite{qiu2024fast}. This problem arises because CASSCF can very often find multiple solutions -- especially for larger active space (but often even for smaller active spaces). Obviously, running dynamics requires smooth potential energy surfaces (at least smooth far away from conical intersections).  To that end, our research group has recently  developed a {\em constrained} CASSCF algorithm \cite{qiu2024fast}, whereby we mandate  that the active space be as small as possible and we further insist that that  active space be partitioned appropriately across a donor and acceptor -- which is appropriate for electron and hole transfer and strongly limits the number of possible solutions. Thus far, at the cost of a more complicated optimization, we have found global potential energy surfaces that are remarkably smooth \cite{qiu2024efficient, qiu2025fast1}. For a chemical example of why a constraint is crucial, see Fig. \ref{fig:intruder}  below.

\subsection{Point of Interest \#2: The Need for an Inexpensive Non-Perturbative SOC Treatment }

For most light-element organic systems, SOC is intrinsically weak. Thus, at or near an equilibrium geometry,
SOC can be treated as a small perturbation to be added on top of the main calculation \cite{fedorov2003spin,marian2021understanding}. 
If the relevant states are well separated,
this perturbation can be treated in a non-degenerate fashion; for example, SOC may only weakly admix
triplet character into a well-separated singlet state.
When several states are nearly degenerate, however, perturbative SOC generally requires
a quasi-degenerate or state-interaction treatment, in which Breit--Pauli spin-dependent
couplings are evaluated among a manifold of spin-free states and the resulting effective
Hamiltonian is rediagonalized \cite{yarkony1992spin, berning2000spin, malmqvist2002restricted,  kahler2023state}.
For instance, if we consider a triplet state and we seek to understand the zero field splitting starting from a spin-free calculation, we need to perturb the small triplet system by all other electronic states  and then rediagonalize. For such a problem, one would need to perturb across a set of many electronic states (which is akin to degenerate perturbation theory).  In short, if we are looking for energy splittings (rather than energy shifts), perturbation theory becomees much more expensive. Finally, we note that such questions will be most apt near curve crossings and conical intersections\cite{matsika2001effects1, matsika2001effects2,matsika2002spin}, where it is known that the presence of a CI can change the topology of a CI\cite{mead:1979:noncrossing, mead:1992:rmp} and where the physics of Berry curvature and molecular phases is critical\cite{resta2000manifestations, wu:2020:neartotal}.
Thus, Point of Interest \#2 regards how can we best and most easily extract non-perturbative information about systems when including SOC. 

On this point, there are two different approaches. Given that our focus in this paper is on electron transfer, all calculations of interest are based on CASSCF simulations that allow for ground-excited state mixing.  One approach is then the  state-interaction treatment referenced above: 
spin-free CASSCF states are computed initially, and SOC is then introduced by constructing and diagonalizing an effective Hamiltonian in
the spin-free CASSCF basis.
This strategy underlies the restricted active space (RAS) state interaction with SOC implementation in
MOLCAS/OpenMolcas~\cite{malmqvist2002restricted, karlstrom2003molcas, aquilante2020modern} and related Quasi Degenerate Perturbation Theory (QDPT) treatments in ORCA \cite{neese2020orca}. Moreover, a more accurate spin-orbit MRCI is also available within the COLUMBUS package\cite{plasser2025columbus, belcher2019analytic, mai2014perturbational}, where numerical gradients are offered. 
Such calculations have been used by Gonz{\'a}lez and co-workers in SHARC simulations of
intersystem crossing (ISC)~\cite{mai2015general,mai2018nonadiabatic}. 
 Note that MRCI calculations can be very expensive, so approximations are always desirable \cite{lischka2018multireference, dombrowski2023r2022}.

An alternative approach is to include SOC directly within CASSCF orbital optimization \cite{ganyushin2013fully}. In other words, while  most CASSCF implementations have been based on real-valued orbitals, one can accommodate SOC directly within CASSCF if one is prepared to work with complex-valued orbitals 
\cite{jenkins2019variational, kasper2020perspective, liao2024comparison, zhang2018spin, wang2026relativistic, shiozaki2018bagel}. 
This latter approach is much less common than the first approach: after all, if one wishes to include  spin-orbit coupling within a CASSCF formalism by including complex orbitals,  one needs not only to change a great deal of code, but convergence is also more difficult given the need for state averaging and the small size of the SOC separating different states.  
Nevertheless, if convergence is possible, then obviously working with this second approach can be computationally efficient; for this direct-SOC approach (as opposed to the state-interaction treatment above), one avoids the need for many response calculations when evaluating the  gradient of the resulting eigenstates during dynamics\cite{gonzalez:2015:ijqc_review}.

In principle, both approaches above are viable. However, in practice, for realistic SOC values, the latter approach (including SOC directly) should converge more quickly with regards to the number of electronic states required. For example,
in their comparison of one- and two-step
SOC treatments within density matrix renormalization group (DMRG), Zhai and Chan showed that many more states are needed for both approaches to agree with each when  SOC mixes a dense low-energy manifold \cite{zhai2022comparison}.
Thus, when deciding which approach to go with, perhaps the most important question is the size of the application of interest. For large-scale applications, MRCI will often not be feasible -- depending on the size of the CI and the cost of the response calculations.  Hence, for our part, where we seek to run large-scale nonadiabatic simulations, we believe the second option will be more computationally efficient  (and we will show below (Sec. \ref{sec:conv_timing}) that the present method is only 10 times slower than a ROHF calculation). 
Admittedly, by not walking down the MRCI route below (and including much more electron correlation), we will sacrifice accuracy. That being said, as we discuss in the next section, our long term goal is qualitative accuracy for spin-dependent ET dynamics, a new area where large scale simulations are desperately needed and one can gain insight without the most chemically accurate electronic structure codes.

\subsection{ Point of Interest \#3: Relevance for Dynamics and the Need of Applicability to a Molecule-Metal Interface Continuum of States}

Although we will not study dynamics here, the question of how nuclei, electrons and spin all couple together during the course of a charge transfer event  is an important area of current research \cite{etherington2016revealing, penfold2018spin, fedorov2018predicting}. For instance, whereas most models of electron transfer are tied to Marcus theory\cite{nitzanbook} in some fashion or another, there is a growing consensus that Marcus theory is incomplete when it  comes to spin-coupled electron transferas is observed through the so-called chiral induced spin selectivity (CISS) effect.\cite{bloom2024chiral}  At bottom, one of the problems with Marcus theory is that the latter does not take into account electronic momentum and conserves the nuclear (rather than total) momentum\cite{bian2023total, tao2024practical, naaman2024can, abedi:2010:prl_exact_factorization, abedi:2012:jcp_exact_factorization}. 
To understand the origin of the problem most simply, consider initially the case without any SOC where an electron transfer process  occurs in the adiabatic limit along the ground state.  (The situation is more complicated for nonadiabatic electron transfer.)
Now, when a molecule vibrates or rotates and the character of the ground state changes, the total electronic momentum (both linear and angular) of the system must shrink or grow.  However, these dynamics are always ignored by  BO  theory. After all, within BO theory, all electronic wavefunctions are real so that $\left< \hat{\bf p}_e \right> = 0$.  In short, BO dynamics conserve the {\em nuclear} dynamics; they do not conserve the total momentum\cite{bian2025phase}. This situation is slightly more complex if we include SOC because now BO states will be complex valued and will include some angular momentum, but momentum conservation will still be completely violated during electron transfer.

To go beyond Marcus theory, and account for both energy conservation {\em and} momentum conservation will require new dynamical techniques that coupled together nuclear, electronic and spin degrees of freedom  -- in addition to new electronic structure techniques. 
Our research group is now actively developing such tools under the general rubric of phase-space electronic structure theory where the electronic Hamiltonian becomes complex-valued and breaks time-reversibility in the electronic space \cite{bian2025phase, peng2026phase}. And within our experience thus far, we have already learned a great deal about how angular momentum conservation rears its head -- for example, in doublet calculations through simulations of the   
diatomic NO molecule. Here,  an experimentally observed lambda doubling splitting arises from an interaction between a pair of degenerate $\Pi$ ground states and a series of much higher level $\Sigma$ electronic states; these states are coupled together through both SOC and a Coriolis coupling.\cite{peng2026conceptual}    Calculations that aim to recapitulate that splitting by first running a spin-free calculation and then performing a perturbative sum can require solving for many excited $\Sigma$ eigenstates besides the ground $\Pi$ states.\cite{hinkley1972lambda}. By contrast, in Ref. \cite{peng2026phase, peng2026conceptual}, we found that a very accurate final result can be found if we initially generate the degenerate ground states directly with SOC and then simply rediagonalize with the Coriolis operator within the SOC included doublet space. 
The end conclusion is that, to correctly capture the effects of angular momentum conservation, one always wants to include SOC as fully and as quickly as possible, which again justifies our interest in small variational CAS calculations with SOC included directly and non-perturbatively.

Finally, given the longer term goal of modeling CISS, it will be very important that we develop methods than can be extended far  beyond charge transfer (with 2 sites). At a minimum, one would like to capacity to study charge transport along a chain of $N$ charge sites efficiently.
More than a discrete model, however, one would also like the capacity to model charge transfer and transport at a metal-molecule interface with a continuum of metallic states. Here, the calculation is far more difficult because one must pick out some electronic states on the molecule and some on the metal if one seeks to model the movement of charge. For this problem, as far as we are aware, the only solution with a CASSCF algorithm is to implement a constraint on the active space\cite{junhan:2023}, as described below. 

\subsection{The Goals of this Paper and an Outline}

\begin{figure}[ht]
    \centering
    \includegraphics[scale=0.7]{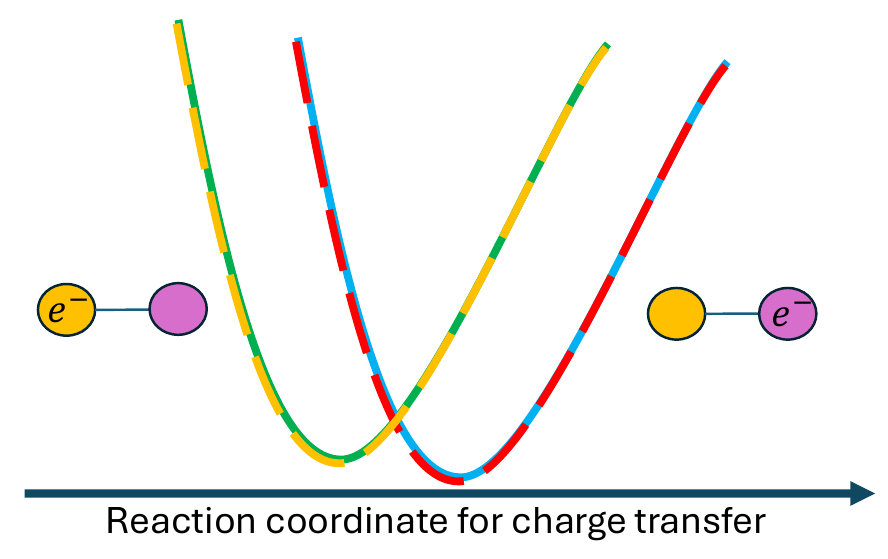}
    \caption{Schematic illustration of crossing between potential energy surfaces (PESs) associated with charge-transfer doublet states in a molecule containing an odd number of electrons. The two diabatic surfaces, correspond to the electron localization of transferring electron on the left or right fragment. The doubly-degenerate nature of each diabat is indicated by colored dashed lines \{yellow,green\} and \{blue, red\}. 
    }
    \label{fig:surfplots}
\end{figure}

Our goal  is to develop the simplest and most efficient electronic structure algorithm based on complex-valued orbitals that can account for SOC directly (and thus will be suitable for phase space electronic structure calculations in the future\cite{bian2025phase}) while also offering globally smooth potential energy surfaces suitable for dynamics -- with possibilities to simulate not just transfer between two charge center sites but also to simulate transport over $N$ charge center sites and a metal-molecule interface. 
With the background above in mind, we will extend the 
Dynamically-weighted State-Averaged Constrained CASSCF (eDSC/hDSC) formalism described in Ref. \cite{qiu2024fast, qiu2024efficient} to the case of electronic states with spin.
Admittedly, small CASSCF simulations are far from standard and will lack a great deal of dynamical correlation, but they are fast and, by including a  constraint on the active space, one can achieve globally smooth surfaces without a dramatic increase in cost. 
As a test case,  we will study a system with odd number of electrons, with reasonably well defined donor/acceptor fragments, and investigate a CT curve crossing between a pair of degenerate Kramers doublets \cite{kramers1930theorie, wigner1993operation}--  
shown heuristically in Fig. \ref{fig:surfplots}. 
Note that the ability to  converge to a very high threshold  is critical if we seek to resolve states with different spin character and very small splittings for future momentum-conserving nonadiabatic dynamics simulations; thus, we will need to carefully check our ability to converge results quickly. While the data presented below is for 2 states, the method has been implemented for  $N$ charge centers as shown in the Appendix (and there have also been previous demonstrations without SOC for a metal-molecule interface\cite{wu2024constrained}).

An outline of this article is as follows. In Sec. \ref{sec:active_space}, we briefly review the constrained eDSC/hDSC theoretical framework and in Sec. \ref{sec:constraint} we formulate the Lagrangian with constraints on the active space. is  In Sec. \ref{sec:spinor}, we define the complex-valued spinor basis in which the Fock and density matrix have different spin blocks; in Sec. \ref{sec:grad} we present the orbital rotation gradients for GHF energy of single Slater determinant; in Sec \ref{sec:full_scf}, we outline the two step orbital optimization algorithm.
The numerical results are presented in Sec. \ref{sec:results}. Sec. \ref{sec:discuss_interuder}  discusses how the imposed constraints helps avoid intruder states. Finally, in Sec. \ref{sec:conclusion}, we conclude our work and discuss possibilities  for future studies of non-adiabatic process and spin chemistry of charge transfer in odd-number electron systems. While the theory in Sec. \ref{sec:method} is written down for simpilicity for the case of 2 electronic states, in the Appendix Sec. \ref{sec:appendix_nstate}, we write down the equations for the case of $N$ electronic states.  As far as notation is concerned, the Greek alphabet will represent the atomic orbitals (AO) and Roman alphabet will represent the molecular orbitals (MO). 

\section{Method}\label{sec:method}

In what follows, we will review and develop eDSC/hDSC for a two-state problem, but we emphasize that our approach can be and has been implemented for an $N$ state approach, as relevant to charge transport (and not just charge transfer). See Appendix \ref{sec:appendix_nstate}. 

\subsection{Review of eDSC/hDSC}\label{sec:active_space}
Our treatment of charge transfer in the presence of an unpaired spin is formulated within a Kramers-restricted open-shell framework.  By enforcing time-reversal symmetry between paired complex molecular orbitals, the construction becomes directly analogous to restricted open-shell Hartree–Fock (ROHF) and corresponds to the “paired GHF” classification in the Stuber–Paldus scheme \cite{stuber2003symmetry, jimenez2011generalized}. The electronic configurations are thus expressed in terms of complex spinors and their time-reversal partners (see Sec. \ref{sec:spinor}), with the resulting wavefunctions naturally occurring in conjugate time-reversal pairs:
\begin{equation}
\begin{aligned}
\ket{\Psi} &= \ket{1\bar{1},\,2\bar{2},\,\ldots,\,N\bar{N},\,N+1},
\quad\quad
\ket{\overline{\Psi}} &= \ket{1\bar{1},\,2\bar{2},\,\ldots,\,N\bar{N},\,\overline{N+1}}
\end{aligned}
\end{equation}
where the orbital pairs, like $1$ and $\bar{1}$, are time-reversal partners.
Following the Dynamically-weighted State-Averaged Constrained CASSCF (eDSC/hDSC) framework of Qiu and Subotnik \cite{qiu2024fast, qiu2024efficient}, the target function to be minimized  is the weighted sum of the energies of these configurations and their time-reversal pairs. The excited configurations are generated through an excitation of an electron (for eDSC) or hole (for hDSC). Because Kramers pairs are degenerate in the absence of time-reversal symmetry breaking (e.g., no external field), it suffices to optimize over a single pair of configurations per state. The total energy is thus given by:
\begin{align}
    E_{\rm tot} &= w_1E_1 + w_2E_2\label{eq:e_tot}
\end{align}

\begin{figure}[ht]
    \centering
    \includegraphics[scale=0.7]{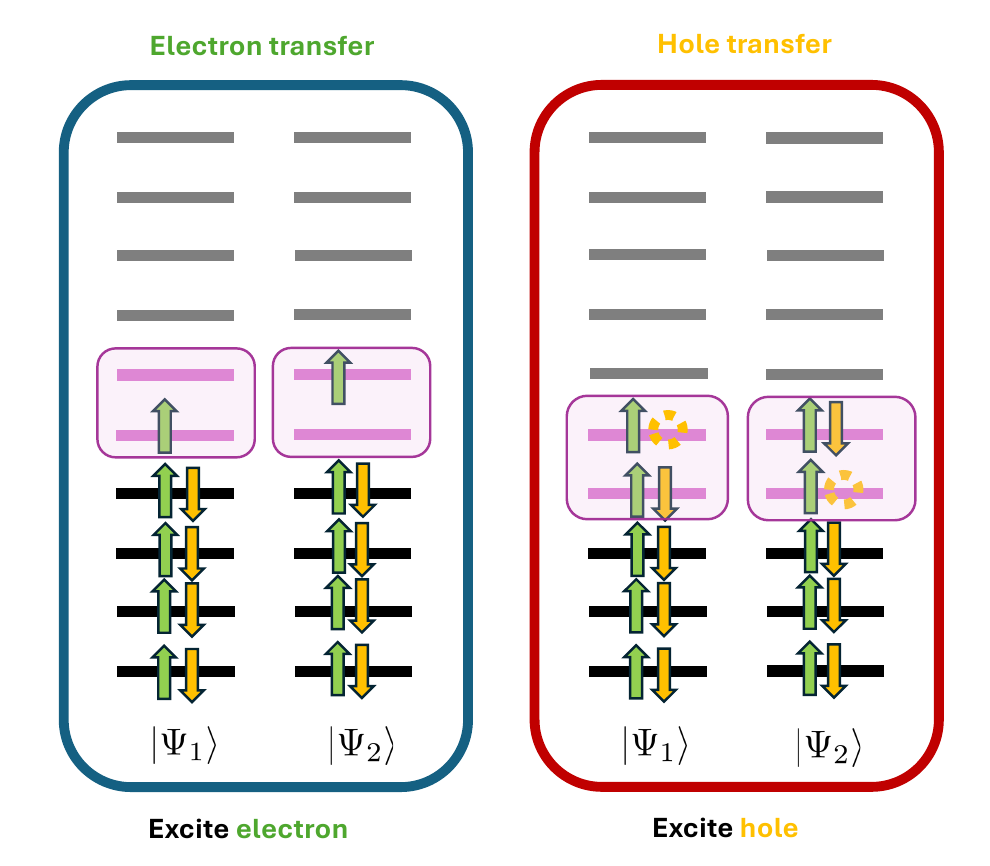}
    \caption{Schematic representation of the active spaces (highlighted in magenta boxes) used to model electron transfer (left) and hole transfer (right) in an open-shell molecular system containing an odd number of electrons. For electron transfer, the active orbitals involve excitation of the unpaired electron, represented by a minimal CASSCF(1,2) active space. The hole-transfer case involves excitation from a doubly occupied orbital, corresponding to a CASSCF(3,2) active space with three electrons distributed over two orbitals.
    }
    \label{fig:cas12}
\end{figure}
For eDSC, the two configurations are
\begin{align}
    \ket{\Psi_1} &= \ket{1\bar{1},2\bar{2},...N\bar{N},N+1}\label{eq:config_e_1}\\
    \ket{\Psi_2} &= \ket{1\bar{1},2\bar{2},...N\bar{N},N+2}\label{eq:config_e_2}
\end{align}
and for hDSC, the two configurations are
\begin{align}
    \ket{\Psi_1} &= \ket{1\bar{1},2\bar{2},...,N\bar{N},N+1}\\
    \ket{\Psi_2} &= \ket{1\bar{1},2\bar{2},...,N,N+1,\overline{N+1}}
\end{align}
The active-space configurations for electron transfer and hole transfer are illustrated schematically in Figure \ref{fig:cas12}.

In Eq. \ref{eq:e_tot}, $w_1$ and $w_2$ are dynamical weights that depend on a ``temperature'' parameter $T$ and the energy difference $\Delta E = E_2 - E_1$, where we assume $E_2 > E_1$. As detailed in the original papers\cite{qiu2024fast,qiu2024efficient}, the weights are proposed to be as follows:
\begin{align}
    w_1(\Delta E) &= 1 - \frac{1-e^{-\Delta E/T}}{2\Delta E/T}\\
    w_2(\Delta E) &= \frac{1-e^{-\Delta E/T}}{2\Delta E/T}
\end{align}
The differential of Eq. \ref{eq:e_tot} is
\begin{align}
    dE_{\rm tot} &= w_1'dE_1 + w_2'dE_2
\end{align}
where
\begin{align}
    w_1' &= (1-\frac{1}{2}e^{-\Delta E/T})\\
    w_2' &= \frac{1}{2}e^{-\Delta E/T}
\end{align}

\subsection{The Essential Need For A Constraint}\label{sec:constraint}
To ensure that the two states correspond to physically meaningful intramolecular charge‐transfer configurations rather than spurious local excitations within individual fragments, a constraint is imposed such that the vector space spanned by the active orbitals (i.e., orbitals $N+1$ and $N+2$ for eDSC and orbitals $N$ and $N+1$ for hDSC) projects equally onto the vector space spanned by orbitals from the donor (or left, $\bm{P}_L$) and acceptor (or right, $\bm{P}_R$) molecular fragments, i.e.,
\begin{align}
    {\rm Tr}(\bm{P}_L\bm{P}_{\rm active} - \bm{P}_R\bm{P}_{\rm active}) = 0\label{eq:q_formal}.
\end{align}

Using a Lagrangian multiplier, one can write the Lagrangian formally as
\begin{align}
    \mathcal{L} = w_1E_1 + w_2E_2 - \lambda {\rm Tr}(\bm{P}_L\bm{P}_{\rm active} - \bm{P}_R\bm{P}_{\rm active}). \label{eq:lag_formal}
\end{align}

\subsection{Spinor Basis: Complex Molecular Orbitals in Generalized Hartree-Fock} \label{sec:spinor}
The presence of spin-orbit coupling term in the electronic Hamiltonian mixes the spin components and spatial degrees of freedom. To properly describe such systems, the formulation requires two-component complex-valued spinors which encode both spin and orbital character. We adopt the Generalized Hartree–Fock (GHF) framework \cite{hammes1993advantages}, which allows for complex orbital coefficients and arbitrary mixing of $\alpha$ and $\beta$ spin components.  
A GHF spin orbital (spinor) can be expanded in terms of real-space atomic orbitals $\phi_{\mu}(\bm{r})$  as 
\begin{equation}
    \chi_i(\mathbf{r}, \omega) = \psi_i^{\alpha}(\mathbf{r}) \, \alpha(\omega) 
                       + \psi_i^{\beta}(\mathbf{r}) \, \beta(\omega),
    \label{eq:ghf_spin_orbital}
\end{equation}
where
\begin{equation}
    \psi_i^{\sigma}(\mathbf{r}) = \sum_{\mu=1}^{K} c_{\mu i}^{\sigma} \, \phi_\mu(\mathbf{r}), \quad \sigma \in \{\alpha, \beta\}
    \label{eq:psi_alpha}
\end{equation}
and $\omega$ is the spin coordinate and $\alpha(\omega)$ and $\beta(\omega)$ are the orthonormal spin functions corresponding to spin-up and spin-down components, respectively. By arranging the molecular orbital (MO) coefficients in column form, the atomic orbital (AO) basis can be organized into a natural spin-block structure, with distinct $\alpha\alpha$, $\alpha\beta$, $\beta\alpha$, and $\beta\beta$ components. In the ordered AO basis, the density matrix and Fock matrix have a four block structure
\begin{equation} \label{eq:spinor_block_PF}
\bm{P} = \begin{pmatrix}
\bm{P}^{\alpha\alpha} & \bm{P}^{\alpha\beta} \\
\bm{P}^{\beta\alpha} & \bm{P}^{\beta\beta}
\end{pmatrix}, \quad
\bm{F} = \begin{pmatrix}
\bm{F}^{\alpha\alpha} & \bm{F}^{\alpha\beta} \\
\bm{F}^{\beta\alpha} & \bm{F}^{\beta\beta}
\end{pmatrix}
\end{equation}

\begin{figure}[ht]
    \centering
    \includegraphics[scale=0.7]{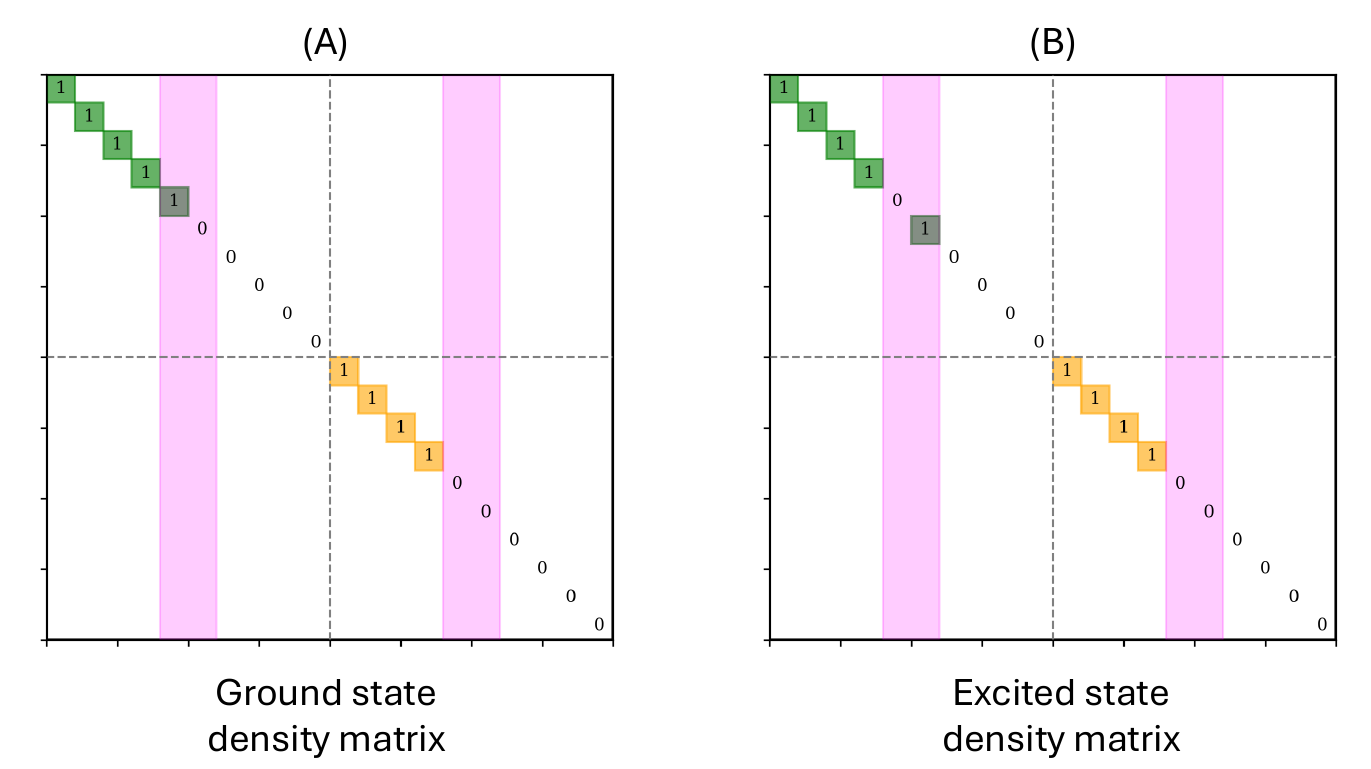}
    \caption{Molecular orbital (MO) density matrices for the (A) ground ($\bm{K_g}$) and (B) excited electronic ($\bm{K_e}$) states. The diagonal elements indicate the occupancies of individual MOs, with pairs of degenerate (Kramers-restricted) orbitals forming two corresponding diagonal sub-blocks (green $(1)$ and yellow $(\bar{1})$). The magenta columns highlight the active-space orbitals involved in the electronic excitation in the spin-up space. Columns to the left (right) of the active space correspond to core (virtual) orbitals in each half of the matrix. 
    }
    \label{fig:Kmats}
\end{figure}

The energies of each configuration, $E_1$ and $E_2$ depend on the density matrices for ground and excited state, respectively,  
\begin{subequations}\label{eq:P_ge}
\begin{align}
    \bm{P}_g &= \bm{C}\bm{K}_g\bm{C}^\dag \label{eq:Pg}\\
    \bm{P}_e &= \bm{C}\bm{K}_e\bm{C}^\dag \label{eq:Pe}
\end{align}
\end{subequations}
which are in turn  expressed in term of occupation matrices 
$\{\bm{K_g}, \bm{K_e}\}$ as visualized in Fig. \ref{fig:Kmats}; here $\bm{C}$ is  the complex molecular orbital coefficient matrix.
The occupation matrix is a diagonal matrix with spin blocks:
\begin{align}
    \bm{K_g} = \bm{K_0} + \bm{\bar{K_0}} + \bm{K_1}\\
    \bm{K_e} = \bm{K_0} + \bm{\bar{K_0}} + \bm{K_2}  
\end{align}
where $\bm{K_0}$ and $\bm{\bar{K_0}}$ are the spinor pairs of the Kramers-restricted core orbitals. $\bm{K_1}$ and $\bm{K_2}$ represent the occupancy of the two active space orbitals. 

The single determinant energy of each configuration state can then be expressed as
\begin{align}
    E[\bm{P}] = \frac{1}{2} {\rm Tr}\left[ (\bm{h}_0 + \bm{h}_{\rm SO} + \bm{F} ) \bm{P}\right]
\end{align}
where $\bm{h}_0$ is the one electron Hamiltonian and $\bm{h}_{\rm SO}$ is the spin-orbit coupling term. The SOC term in Breit--Pauli Hamiltonian contains a one electron part (term arising from the interaction of moving electrons with nuclei) and a two-electron part (made of spin-same-orbit and spin-other orbit contribution) \cite{fedorov2000study}. The one-electron part, being local in nature, is computationally inexpensive to calculate. That being said, it is known that including the two-electron term can decrease the overall magnitude of SOC\cite{boettger2000approximate}. As such, there are various schemes based on effective one-electron SOC term which account for the  contribution from the two-electron part using a mean-field operator \cite{hess1996mean, tatchen1999performance, liu2018atomic}, an effective nuclear charge \cite{koseki1995main} or other screening effects\cite{ehrman2023improving}. For the sake of simplicity, here we ignore screening and consider the simple bare one-electron term in the Breit-Pauli Hamiltonian:
\begin{align} \label{eq:h_so}
    \bm{\hat{h}}_{\rm SO}(\bm{r}) = \frac{\alpha_{0}^{2}}{2} \quad \sum_{i=1}^{N_{elec}} \sum_{N=1}^{N_{nuc}} \frac{Z_N}{|\bm{\hat{r}}_i-\bm{\hat{R}}_N|^3} \left( (\bm{\hat{r}}_i-\bm{\hat{R}}_N) \times \bm{\hat{p}}_i \right)\cdot\bm{\hat{s}}_i
\end{align}
where $\alpha_0$ is the fine structure constant. Indexes $i$ and $N$ label electrons and nuclei, respectively, with $\bm{\hat{r}}_i$ and $\bm{\hat{R}}_N$ being their corresponding position operators. $\bm{\hat{p}}_i$ and $\bm{\hat{s}}_i$ are linear momentum and spin operators, respectively, for electron $i$. $Z_N$ is the charge of nucleus N.
In principle, we should of course include not just the one-electron SOC in Eq. \ref{eq:h_so} but also the corresponding two-electron SOC (or the mean-field approximation thereof\cite{marian:2001:revcompchem, hess1996mean}). For instance, it is known that the latter usually reduces  the strength of the former, often by 50\%. But given our desire for qualitative accuracy, Eq. \ref{eq:h_so} should suffice for now.

The matrix representation of the spin–orbit Hamiltonian term in the atomic orbital basis contains elements proportional to the scalar product 
between electronic orbital angular momentum ($\bm{L_{elec}}$) and spin ($\bm{S}$) angular momentum.
This leads to a block structure as in Eq. \ref{eq:spinor_block_PF}, with off-diagonal blocks introducing complex-valued terms into the electronic Hamiltonian:
\begin{align}
    \bm{L_{\rm elec}}\cdot\bm{S} 
    & \sim
    \begin{pmatrix}
    L_{\mu\nu}^z                & L_{\mu\nu}^x - i\,L_{\mu\nu}^y \\
    L_{\mu\nu}^x + i\,L_{\mu\nu}^y & -\,L_{\mu\nu}^z
    \end{pmatrix}
\end{align}
Finally we have the constraint  as defined in Eq. \ref{eq:q_formal} in terms of projections onto the active space,
\begin{align}
     {\rm Tr}[\bm{Q}(\bm{P_1}+\bm{P_2})] = 0
\end{align}
where $\bm{Q}=\bm{P_L}-\bm{P_R}$,
and active space densities $\bm{P_1} = \bm{C} \bm{K_1} \bm{C^{\dagger}}$ and $\bm{P_2} = \bm{C} \bm{K_2} \bm{C^{\dagger}}$. $\bm{P_L}$ and $\bm{P_R}$ do not involve spin and have the identical spin structure.

\subsection{Electronic Gradients for Energy and Constraints}\label{sec:grad}
To variationally minimize the target function subject to spin symmetry and active space projection constraints (Eq. \ref{eq:lag_formal}), we optimize the molecular orbitals via a unitary transformation of the initial guess of coefficient matrix $\bm{C}_0$.
We define an anti-Hermitian orbital rotation matrix $\bm{A}$ in the molecular orbital (MO) basis and construct the unitary transformation $\bm{U} = \bm{e^A}$. The orbitals are updated using this rotation matrix 
\begin{align} \label{eq:C_U}
    \bm{C} = \bm{C_0} \bm{e^A}
\end{align}
The gradient of the GHF energy (say, in the ground state with density matrix $\bm{K}_g$) is then (more details in Appendix \ref{sec:appendix})
\begin{align}
    \nabla_{A_{pq}}E = 
    \left (\left[\bm{(\bm{C_0}^\dagger\bm{F}\bm{C_0})^{\top}, \bm{K_g}} \right] \right )_{pq}
    \label{eq:grad_Apq}
\end{align}
Under time-reversal symmetry, MO coefficients transform as follows:
\begin{equation} \label{eq:trs_C}
\hat{T}
\begin{pmatrix}
c^{\alpha}_{\mu i} \\
c^{\beta}_{\mu i}
\end{pmatrix}
=
\begin{pmatrix}
-c^{\beta*}_{\mu i} \\
c^{\alpha*}_{\mu i}
\end{pmatrix}
\end{equation} 
To ensure that any symmetry present in the initial guess density matrix is preserved throughout the orbital optimization, the same symmetry constraints must be imposed on the orbital rotation matrix. To that end, note that unitarity is guaranteed for any transformation of the MO coefficients with an anti-Hermitian  generator matrix $\bm{A}$.
Moreover, to preserve time-reversal symmetry between the molecular orbitals, 
$\bm{A}$ must adopt a certain structure:
\begin{align} \label{eq:trs_block}
    \bm{A} = \begin{bmatrix}
        \bm{B} & -\bm{D}^* \\
        \bm{D} & \bm{B}^*
        \end{bmatrix}   
\end{align}
Note that this form eliminates any chance of artificial time-reversal symmetry breaking, which can be extremely problematic for GHF calculations involving SOC.\cite{kasper2020perspective}

The gradient of the constraint has a similar form as the commutator in Eq. \ref{eq:grad_Apq}. Thus, the full expression for the gradient of the Lagrangian  becomes 
\begin{align} 
    \nabla_{A_{pq}}\mathcal{L}
    &= w_1' \nabla_{A_{pq}}{E_1} + w_2' \nabla_{A_{pq}}{E_2}  - \lambda \nabla_{A_{pq}} {\rm Tr}[\bm{Q}(\bm{P}_1+\bm{P}_2)] \\
    &= w_1' [\bm{M}_1^{\top},\bm{K}_g] + w_2' [\bm{M}_2^{\top},\bm{K}_e] - \lambda [\bm{M}_Q^{\top},\bm{K}_1 +\bm{K}_2] 
    \label{eq:lag_scf}
\end{align}
where
\begin{align}\label{eq:M0}
    \bm{M}_1 &= \bm{C_0}^{\dagger} ( \bm{F}[\bm{P}_g] )  \bm{C_0} \\ 
    \bm{M}_2 &= \bm{C_0}^{\dagger} ( \bm{F}[\bm{P}_e] )  \bm{C_0} \\ 
    \bm{M}_Q &= \bm{C_0}^{\dagger} \bm{Q}  \bm{C_0} \label{eq:MQ}
\end{align}

When interpreting the gradient of the Lagrangian in Eq. \ref{eq:lag_scf}, several nuances must be pointed out.
First, note that each block of the rotation matrix has real and imaginary parts.
Thus, the gradient  in Eq. \ref{eq:lag_scf} must be taken w.r.t. the real and imaginary part of $\bm{B}$ ($\bm{B}^r,\bm{B}^i$) and $\bm{D}$ ($\bm{D}^r,\bm{D}^i$) in Eq. \ref{eq:trs_block}.
Second, in order to maintain orthonormality of orbitals and preserve time-reversal symmetry during orbital optimization, the $\bm{B}$ block should be anti-Hermitian, so that  $\bm{B}^r$ should be anti-symmetric and $\bm{B}^i$ symmetric. Similarly, $\bm{D}$ (and hence $\bm{D}^r$ and $\bm{D}^i$) should be symmetric. We capture all these symmetries mathematically in Eq. \ref{eq:grad_BD}.
Third, consider the diagonal elements of the $\bm{B}$ and $\bm{D}$ blocks. The diagonal elements of $\bm{B}$ corresponds to a phase change; a diagonal element of $\bm{D}$ corresponds to mixing one state with its time reversal pair, which will be degenerate if the Hamiltonian has time-reversal symmetry.
Consequently, we exclude these diagonal terms from the gradient evaluation.
Fourth,  note that only inter-subspace rotations within the occupied-unoccupied subspace of the  $\bm{B}$ and $\bm{D}$ blocks  contribute to a change in the energy and hence only those elements in the upper triangle need to be included in the gradient calculation. Let  us now  be more precise mathematically. Let us divide up $\nabla_{A_{pq}}\mathcal{L} \equiv \bm{G}$ into four blocks as
\begin{align}
    \bm{G} = \begin{pmatrix}
        \bm{G}^{11} &\bm{G}^{12}\\
        \bm{G}^{21} &\bm{G}^{22}
    \end{pmatrix}.
\end{align}

As discussed above,
imposing time-reversal structure on the rotation matrix means that we have to take gradients w.r.t. real and imaginary parts of $\bm{B}$ and $\bm{D}$ matrices.
\begin{subequations}\label{eq:grad_BD}
\begin{align}
    \nabla_{\bm{B}^r}\mathcal{L} &= (\bm{G}^{11}-\bm{G}^{11\top})+(\bm{G}^{22}-\bm{G}^{22\top}), \label{eq:grad_BDr}\\
    \nabla_{\bm{B}^i}\mathcal{L} &= i\left((\bm{G}^{11}+\bm{G}^{11\top})-(\bm{G}^{22}+\bm{G}^{22\top})\right), \label{eq:grad_BDi}\\
    \nabla_{\bm{D}^r}\mathcal{L} &= (\bm{G}^{21}+\bm{G}^{21\top})-(\bm{G}^{12}+\bm{G}^{12\top}), \label{eq:grad_Dr}\\
    \nabla_{\bm{D}^i}\mathcal{L} &= i\left((\bm{G}^{21}+\bm{G}^{21\top})+(\bm{G}^{12}+\bm{G}^{12\top})\right). \label{eq:grad_Di}
\end{align}
\end{subequations}

Note that all four gradients above are real-valued functions. Solving a constrainede optimization problem as in (Eq. \ref{eq:lag_formal}), is usually done using a SQP \cite{Nocedal2006} (Sequential Quadratic Programming) algorithm which requires only the gradients of the target function (Eq. \ref{eq:e_tot}) and the constraint (Eq. \ref{eq:q_formal}). In the spirit of  Ref. \citenum{qiu2024efficient}, we will initiate the orbital optimization through a two step iterative procedure: $(i)$ a SCF step  and $(ii)$ a constrained non-self-consistent-field (nSCF) step.

\subsubsection{SCF steps}

In the SCF steps, the complex MO coefficients are updated by solving the Lagrangian in Eq. \ref{eq:lag_formal}.  A further speedup is obtained when using the direct inversion of iterative subspace (DIIS) technique to solve the SCF problem \cite{pulay1980convergence, pulay1982improved}. The gradient of Lagrangian should vanish at convergence. Thus Eq. \ref{eq:lag_scf}, suggests that we can use the following form of error vector in DIIS:
\begin{align} \label{eq:V_diis}
    \bm{V}_{\rm DIIS} 
    &= w_1' [\bm{M}_1^{\top},\bm{K}_g] + w_2' [\bm{M}_2^{\top},\bm{K}_e] - \lambda [\bm{M}_Q^{\top},\bm{K}_1 +\bm{K}_2]
\end{align}
indicating how far the electronic densities are from the true solution.
Again, in order to maintain the symmetries,  the actual DIIS vector would be
\begin{align} \label{eq:diis_sym}
    \bm{V}_{\rm DIIS}  = \begin{pmatrix}  \grad{\Br}\mathcal{L} \\  \grad{\Bi}\mathcal{L} \\  \grad{\Dr}\mathcal{L} \\ \grad{\Di}\mathcal{L} \end{pmatrix}.
\end{align}

\subsubsection{nSCF steps}
For the nSCF step, we solve for new MO coefficients using fixed values of $w_1^{'} , w_2^{'}$ and $\bm{\tilde{F}}$. Here, $\bm{\tilde{F}}$ ($\bm{\tilde{F}_g}$ and $\bm{\tilde{F}_e}$ for ground and excited state, respectively) is the Fock matrix calculated for a fixed value of the density matrix.  We define an auxiliary energy function of the form:
\begin{align}
    E_{\rm aux} = w_1^{'}{\rm Tr}[\bm{\tilde{F}_g}\bm{P_g}] + w_2^{'}{\rm Tr}[\bm{\tilde{F}_e}\bm{P_e}].  
\end{align}
Minimizing $E_{\rm aux}$ with the constraint on active space is equivalent to solving a Lagrangian of form
\begin{align}\label{eq:lag_aux}
    \mathcal{L}_{\rm aux} = w_1^{'}{\rm Tr}[\bm{\tilde{F}_g}\bm{P_g}] + w_2^{'}{\rm Tr}[\bm{\tilde{F}_e}\bm{P_e}]   - \lambda {\rm Tr}(\bm{Q} (\bm{P}_1 + \bm{P}_2)). 
\end{align}

\subsection{Full SCF procedure}\label{sec:full_scf}
Finally, let us summarize the proposed procedure for calculating the energy. The idea is to solve the SCF equations to find the optimal set of orbital rotation matrices under some constraints. We work in the complex spinor basis which has a block structure as shown in Eq. \ref{eq:spinor_block_PF}. To enforce this block structure and maintain time reversal symmetry (TRS), we diagonalize the spin up block of the core hamiltonian and use the eigenvectors to construct orthonormal MO coeffients as Kramers restricted pairs (Eq. \ref{eq:trs_C}). Since the updates to this MO coefficient matrix occur through unitary rotation,
the TRS relation is maintained in each SCF update (Eq. \ref{eq:C_U}) until convergence.

The algorithmic steps are as follows:
\begin{enumerate}
    \item Choose a set of MO coefficients  ($\bm{C}_{\text{ini}}$); calculate $\bm{Q}$.
    \item Build ground ($\bm{P}_g$) and excited ($\bm{P}_e$)  state density matrices using Eq. \ref{eq:P_ge}
    \item Build state weights ($w_1$ and $w_2$) and Fock matrices ($\bm{F[P_g]}$ and $\bm{F[P_e]}$) using density matrices . Calculate Fock matrices and $\bm{Q}$ matrix in MO basis using Eq. \ref{eq:M0} - \ref{eq:MQ}. \label{{step:diis_start}}
    \item Compute the initial DIIS error vector $\bm{V}^0$ using Eq. \ref{eq:V_diis}, i.e.,
    \begin{align}
        \bm{V}^0 
        &= w_1' [\bm{M}_1^\top,\bm{K}_g] + w_2' [\bm{M}_2^\top,\bm{K}_e] - \lambda [\bm{M}_Q^\top,\bm{K}_1 +\bm{K}_2] 
    \end{align}  
    and impose proper symmetries as in Eq. \ref{eq:diis_sym}.

\item If norm $(\bm{V}^n) < T_D$ (DIIS error threshold), jump to {\bf Step} \ref{step:diis_stop}. Otherwise, set the initial DIIS variable $\bm{A}^0$ (Eq. \ref{eq:diis_var}) to $\bm{0}$.
    \item Solve the nSCF problem for the new $\bm{C}$ and $\lambda$ using SQP. The nSCF threshold for the norm of the gradient and the constraint is set to norm $(\bm{V}^0)/100$.
    \item (DIIS iteration starts) Build density matrices from new $\bm{C}$ using Eq. \ref{eq:P_ge}
    \item Build state weights and calculate Fock and $\bm{Q}$ matrices in MO basis
    \item Calculate the DIIS error   vector $\bm{V^n}$ at iteration $n$.
    \item Solve the nSCF problem for new $\bm{C}$ and $\lambda$ using SQP.
    The nSCF threshold for the norm of the gradient and the constraint is set to norm $(\bm{V}^n)/100$.\label{step:nscf}
    \item Calculate the DIIS variable at iteration $n$:
    \begin{align} \label{eq:diis_var}
        \bm{A}^n = \log\left(\bm{C}_{\rm ini}^\dagger\bm{C}\right).
    \end{align}
    \item If norm $(\bm{V}^n) < T_D$, exit the iteration and jump to {\bf Step} \ref{step:diis_stop}. Otherwise, use DIIS algorithm to calculate new $\bm{C}$. 
    and set $\bm{C}$ to be $\bm{C}_{\rm ini}e^{\Delta \bm{A}}$.
    \item Return to {\bf Step} 7 
    \item DIIS convergence achieved.\label{step:diis_stop}
\end{enumerate}

\section{Numerical Results}\label{sec:results}
\begin{figure}[ht]
    \centering
    \includegraphics[width=0.9\textwidth]{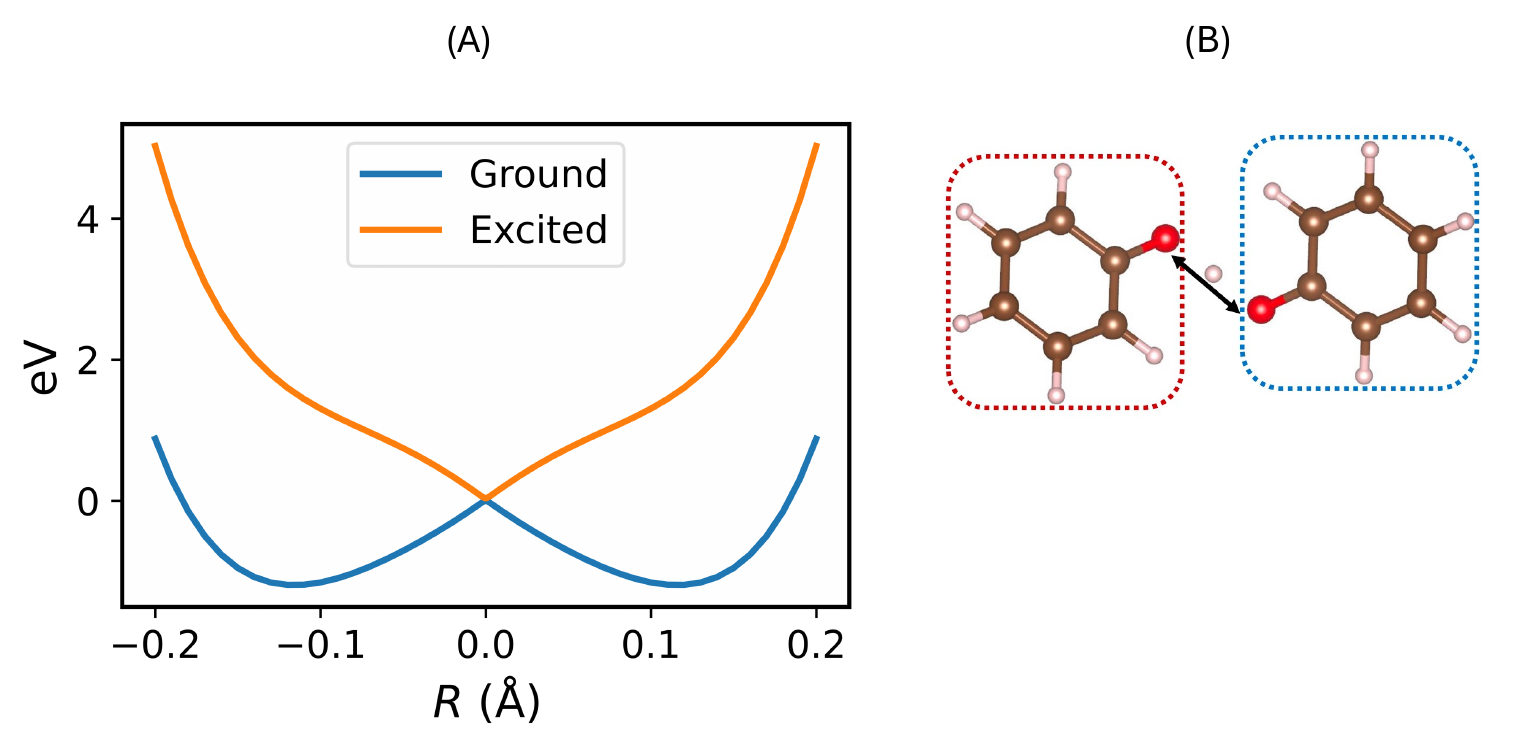}
    \caption{
    Potential energy curves for a bridging hydrogen transferring between two symmetrically located fragments. (A) Ground and excited energy surfaces shown in blue and orange, respectively. (B) Molecular geometry of the phenoxyl–phenol (ph–ph) system, illustrating the hydrogen transfer pathway (black arrow) between the two ring fragments highlighted in red and blue. The reaction coordinate $R$ represents the displacement of the transferring hydrogen relative to the midpoint between the two oxygen atoms. The internuclear distance between oxygen atoms is 2.459 \AA. See Supplementary Information for Cartesian coordinates of molecular geometry.}
    \label{fig:pes_adiab}
\end{figure}

We have implemented the  algorithm described above in a development version of Q-Chem \cite{epifanovsky2021software}  and applied it to a charge transfer system.
The phenoxy-phenol (ph-ph) system has been studied before within a restricted open-shell framework using hDSC \cite{qiu2024fast, qiu2024efficient}.
The result (in  the presence of SOC) is presented in 
Fig. \ref{fig:pes_adiab}, where the reaction coordinate is the displacement of bridging hydrogen between oxygen atoms on symmetrically located fragments. Note that each curve is doubly degenerate, representing the symmetry between time-reversal configuration pairs in the ground and excited states.  As established previously, we find a very natural curve crossing here as the hydrogen position is correlated with the position of the electron, i.e., this model represents a proton-coupled electron transfer (PCET)\cite{hammes2015proton} process.

\subsection{Rate of Convergence and Timings} \label{sec:conv_timing}

To assess the convergence of the two-step SCF procedure outlined in Sec.~\ref{sec:full_scf}, we monitor the decay of the DIIS error-vector norm, $\|\mathbf{V}^{n}\|$, as a function of the outer DIIS-SCF iteration. Since $\mathbf{V}^{n}$ is the residual gradient of the Lagrangian in Eq. \ref{eq:lag_scf}, this norm provides a direct measure of convergence for the constrained SCF stationarity condition. The second step of the algorithm, in which the auxiliary Lagrangian in Eq.~\ref{eq:lag_aux} is minimized through constrained nSCF/SQP iterations, is relatively inexpensive because the Fock matrices are kept fixed. As a result, these inner nSCF iterations avoid repeated Fock builds, which is otherwise a major computational bottleneck in the complex spinor basis.

\begin{figure}[ht]
    \centering
    \includegraphics[width=0.6\textwidth]{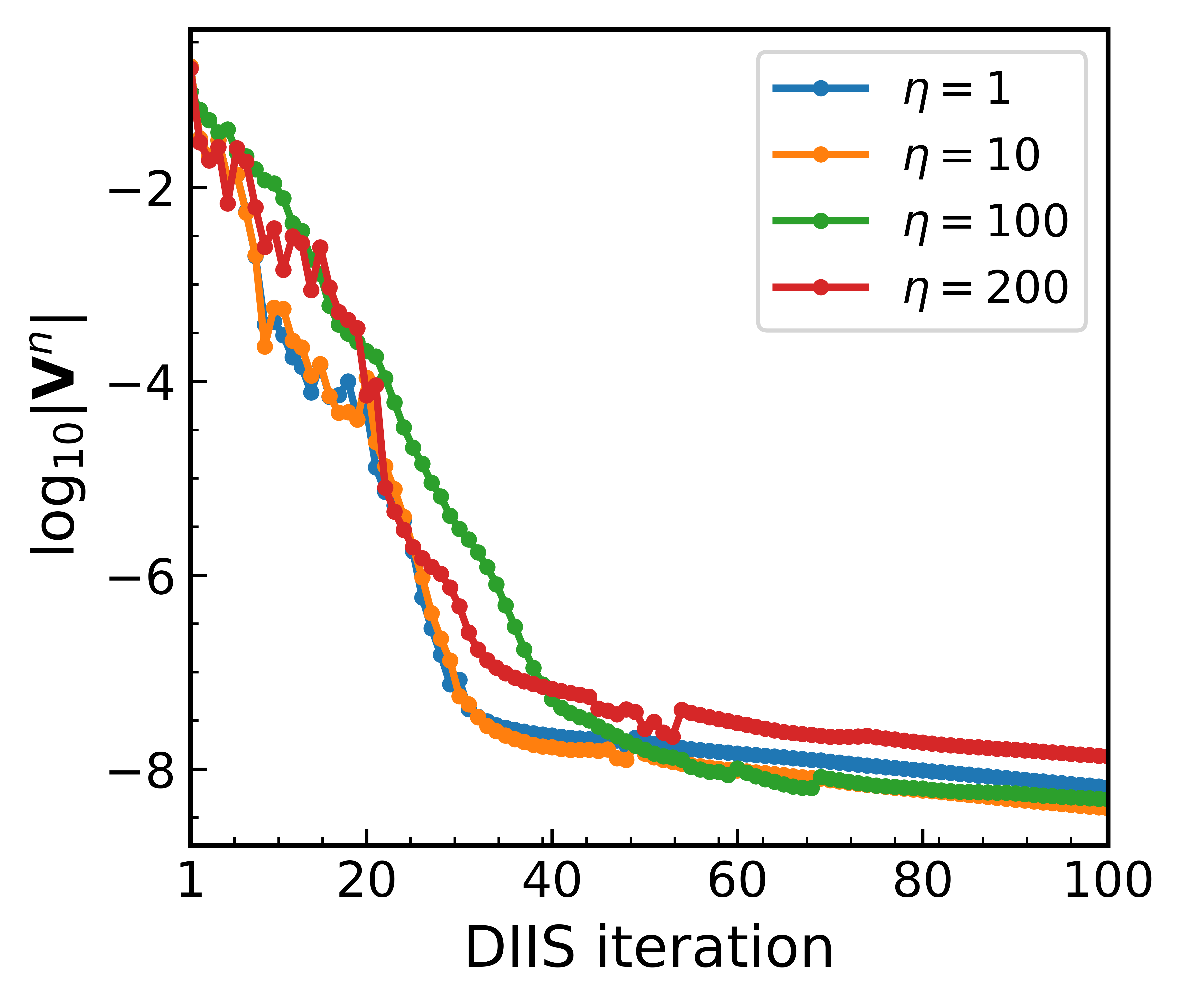}
    \caption{
    DIIS convergence for eDSC/hDSC SCF solver. The residual norm $\rm log_{10} |\bm{V}^n|$ decays rapidly for different SOC strengths $\eta$ = [1,10,100,200], indicating robust convergence from weak to strongly scaled SOC regimes.}
    \label{fig:diis_conv}
\end{figure}

Figure~\ref{fig:diis_conv} shows the corresponding convergence behavior for several values of the SOC scaling parameter $\eta$. In all cases, the DIIS error norm decreases rapidly over the first few tens of outer DIIS-SCF iterations, demonstrating that the two-step SCF procedure remains stable over a broad range of SOC strengths.  These results indicate that the constrained complex-spinor SCF algorithm is robust not only for the physical case (where the SOC is weak), but also for artificially enhanced SOC values that strongly mix spin and orbital character.

\begin{figure}[H]
    \centering
    \includegraphics[width=0.6\textwidth]{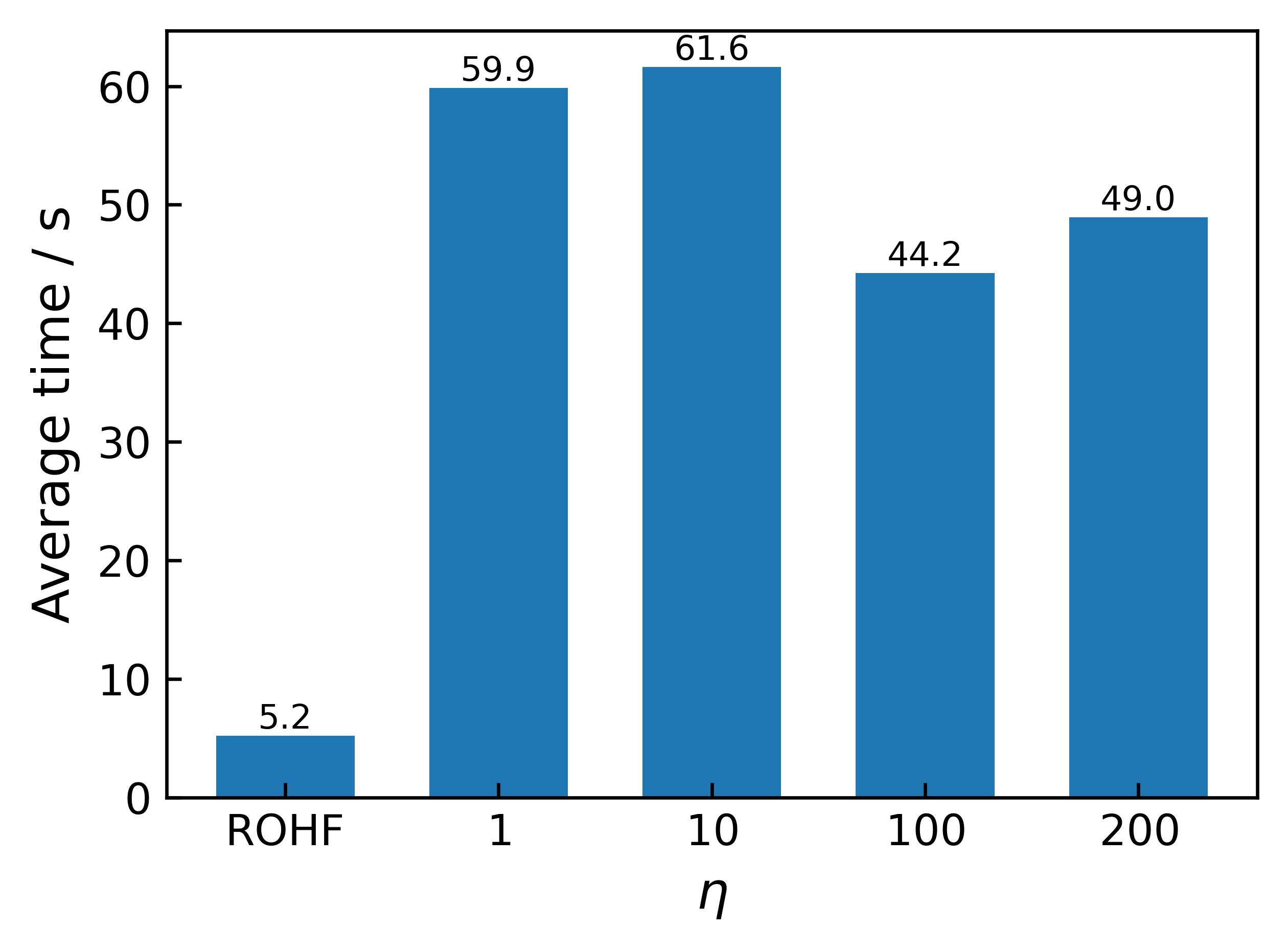}
    \caption{
    Average SCF wall times along the reaction coordinate. The ROHF timing is shown as a reference, followed by eDSC/hDSC DIIS-SCF timings for different SOC strengths. These calculations were performed on 8 cores.}
    \label{fig:diis_conv_time}
\end{figure}

In Fig. \ref{fig:diis_conv_time}, we further plot the total wall time for our calculations as a function of $\eta$. Note that our calculations are  efficient, only a factor of 10 relative to an ROHF calculation without SOC, indicating suitability for future nonadiabatic simulations. (Note that, one should to lose a factor of 3 or so when moving from real to complex orbitals by including SOC; one should also expect to lose a factor of 2 when one solves for ground+excited state [and not just the ground state alone]; thus the best one could hope for would be a factor of 6.)   To generate the data in Fig. \ref{fig:diis_conv_time}, we have scanned from left to right on the reaction coordinate (-0.2 to 0.2 \AA) and use the converged solution at previous geometry as a guess for the next one. For a fair comparison, we exclude the timing for the first geometry from both the ROHF and DIIS-SCF benchmarks. For the ROHF calculations, the standard DIIS optimizer in Q-Chem failed to reduce the DIIS error below $10^{-5}$ for some geometries near the crossing. We had to use the DIIS-GDM optimizer to be able to convergence below $10^{-9}$ Hartree in energy. The ROHF timings reported in Fig.~\ref{fig:diis_conv_time} correspond to these DIIS-GDM calculations.

\subsection{Scaling spin-orbit coupling strength}
\begin{figure}[ht]
    \centering
    \includegraphics[scale=0.55]{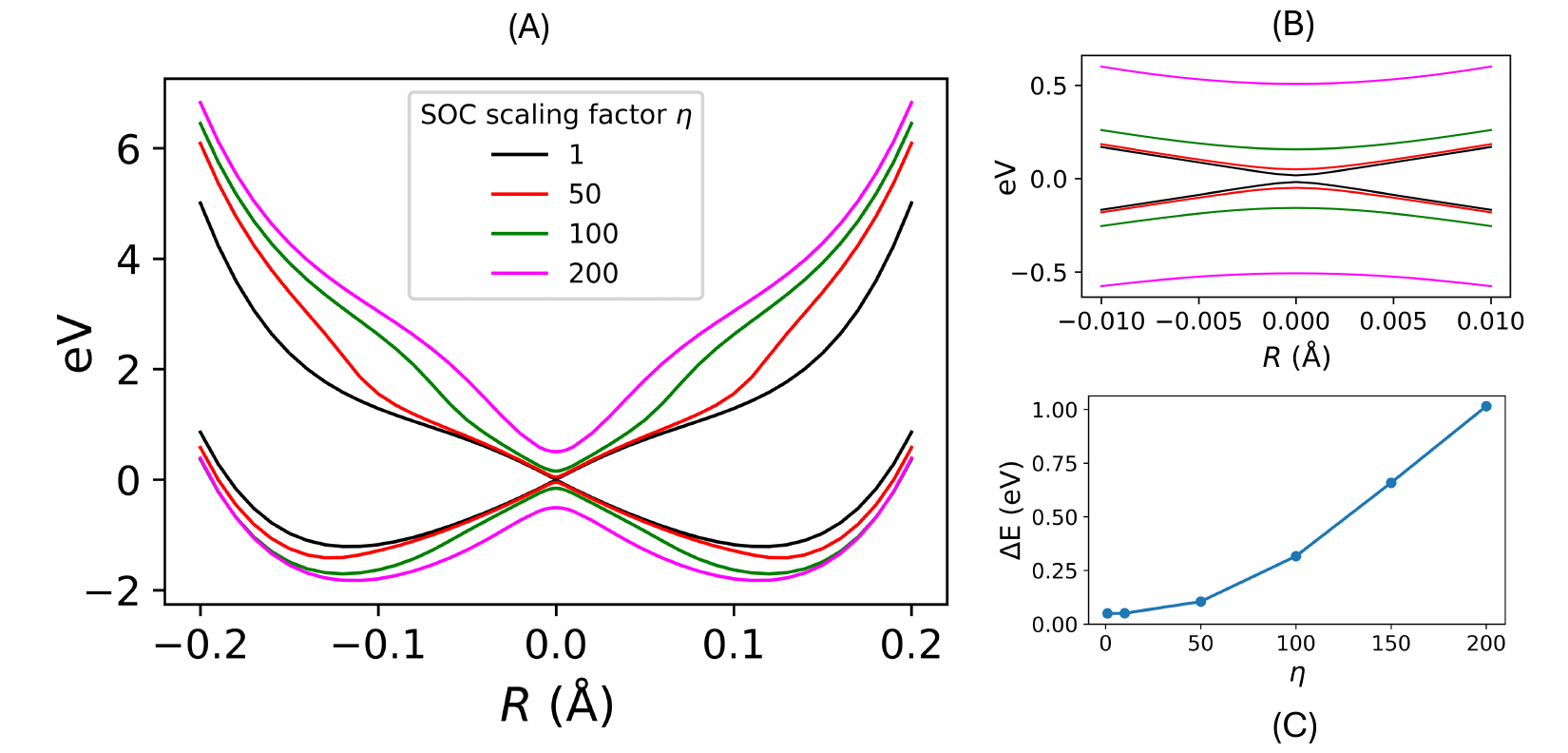}
    \caption{Effect of increasing spin-orbit coupling strength. (A) Ground and excited state PES. The energies are shifted to set zero as mean of two surfaces in the middle. The SOC term in Eq. \ref{eq:h_so} is scaled by a factor $\eta$ for all geometries along the reaction coordinate. Each color represents the two PES for $\eta$ = [1, 50, 100, 200]. 
    (B) PES near avoided crossing region. (C) The energy gap between ground and excited state at $R = 0.0\ \text{\AA}$ for different values of $\eta$.
    }\label{fig:socX}
\end{figure}

Figure \ref{fig:pes_adiab}
shows that, even in presence of SOC, the potential energy curves remain qualitatively similar to the spin-free results reported in Ref. \cite{qiu2024fast}. We next use this system to ask a different question: how do the potential energy surfaces and the crossing region respond as the strength of spin--orbit coupling is artifically increased? Our goal here is not to obtain quantitatively accurate SOC contributions for a specific molecule, as would be required for direct comparison with spectroscopy, but rather to use SOC scaling as a controlled probe of the general trends near charge-transfer crossings and  to assess the stability or our algorithm to the larger perturbations that will arise for other molecules and atoms.
To this end, we introduce a dimensionless scaling parameter $\eta$, which multiplies the one-electron Breit–Pauli SOC term in Eq. \ref{eq:h_so}, and compute the corresponding energy surfaces for several values of $\eta$. The result is shown in Fig. \ref{fig:socX} (A), with energy gap near avoided crossing plotted in Panel (B) and (C).

Two features emerge from these figures. First,  the general shape of the curve can change with $\eta$, but the algorithm appears robust and our curves remain smooth.  Second and most importantly, we correctly find that the energy gap in the central crossing region (i.e. the diabatic coupling) increases with $\eta$.  Panel B zooms in the avoided crossing region of the PES for these scaling parameters.  Panel C highlights the fact that the energy gap scales quadratically, as one should expect.

\subsection{Electron density in active orbitals}

\begin{figure}[ht]
    \centering

    % ---------- (a) ----------
    \begin{subfigure}{\textwidth}
        \centering
        \includegraphics[scale=0.55]{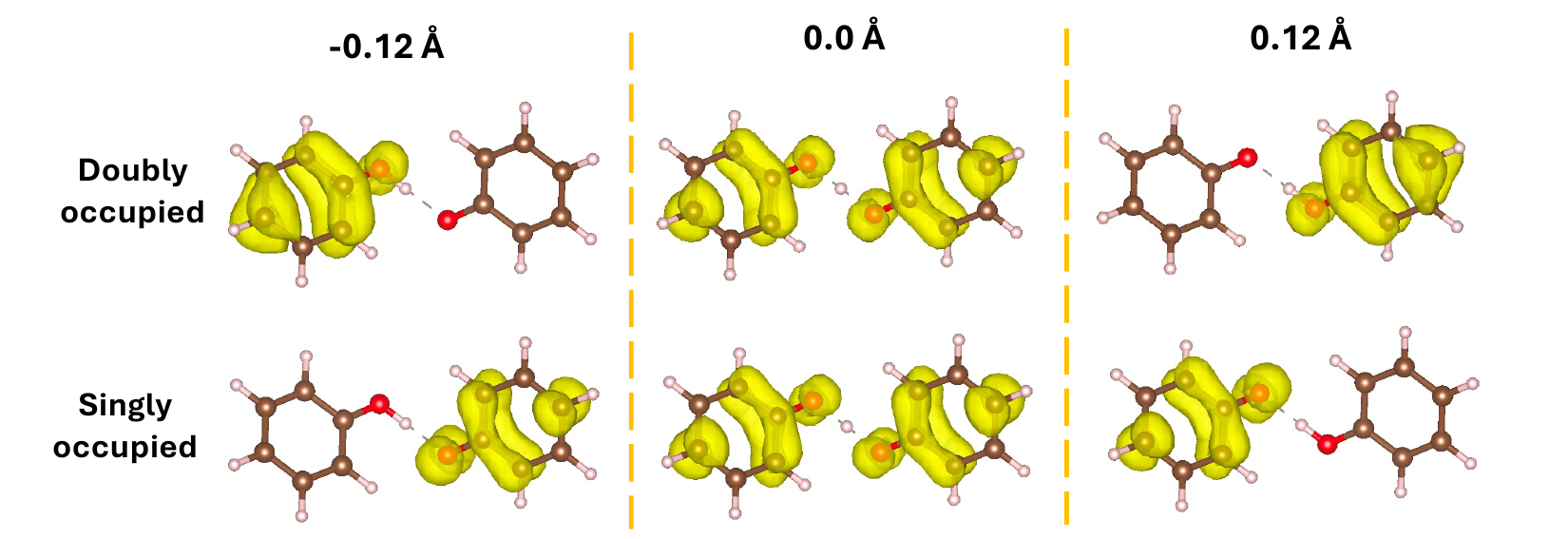}
        \caption{Active orbital electron density with SOC strength $\eta = 1$.}
        \label{fig:cube_soc1}
    \end{subfigure}

    \vspace{1em} % space between the panels 

    % ---------- (b) ----------
    \begin{subfigure}{\textwidth}
        \centering
        \includegraphics[scale=0.55]{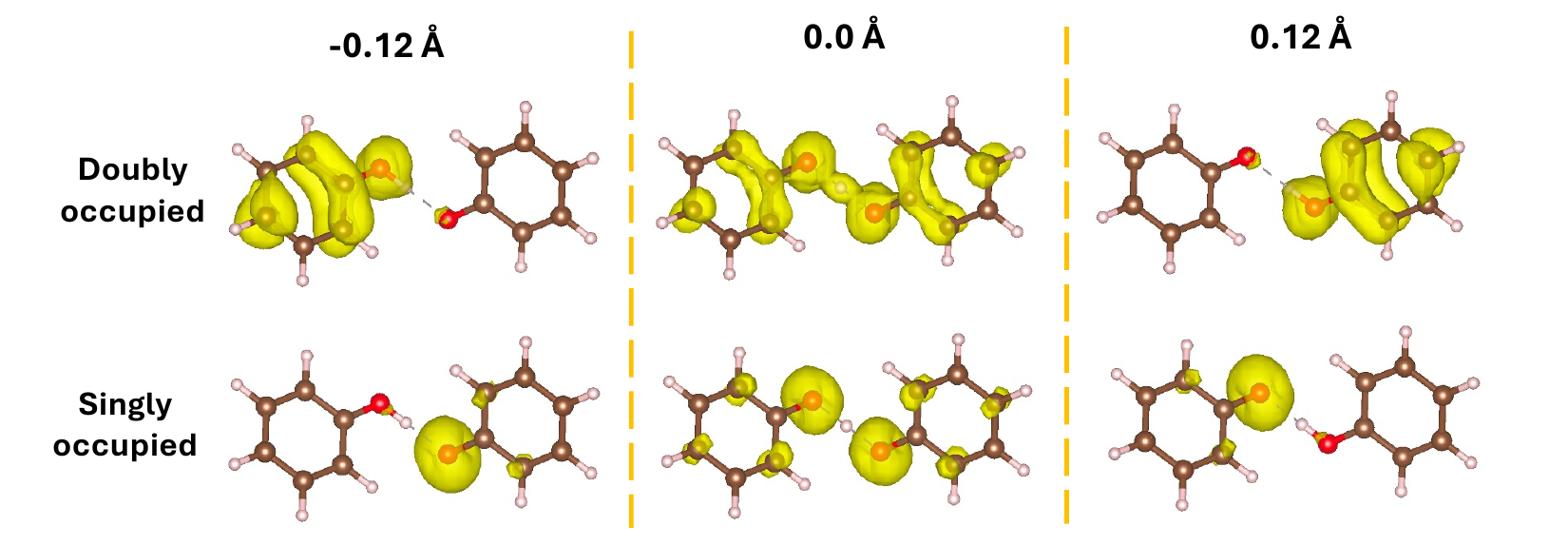}
        \caption{Active orbital electron density with SOC strength $\eta = 200$.}
        \label{fig:cube_soc200}
    \end{subfigure}

    \caption{A comparison of the active orbital electron densities for two different SOC coupling strengths (A) $\eta = 1$ and (B) $\eta = 200$ at R = [-0.12 \AA, 0.0 \AA, 0.12 \AA] }
    \label{fig:cube_soc_combined}
\end{figure}

In real space, the electronic density at position $\br_0$ is the expectation value of the operator $\hat{\rho}(\br_0) = \delta(\hat{\br} - \br_0) = \ket{\br_0}\bra{\br_0}$. Given 
 the usual occupied-orbital expansion, the total  real-space electron density can be calculated as:
\begin{align}
    \rho(\mathbf r)
    &= \left< \Psi \middle| \br \right>\left<\br \middle| \Psi\right>
    \\ 
    &= \sum_{i \in \text{occ}}
      \left< \chi_i \middle| \br\right> \left< \br \middle| \chi_i\right> 
    \label{eq:rho_r_def}
    \\
    &= \sum_{i \in \text{occ}}
      \left< \chi_i \middle| \br_{\alpha}\right> \left< \br_{\alpha} \middle| \chi_i\right>  + \left< \chi_i \middle| \br_{\beta}\right> \left< \br_{\beta} \middle| \chi_i\right>
\end{align}
where we have inserted the spin identity for a two-level system in the last row:
$\mathcal{I} = \ket{\alpha}\bra{\alpha} + \ket{\beta}\bra{\beta}$.
If we expand in terms of the spin orbitals from Eq. \ref{eq:ghf_spin_orbital}, the electron density of a GHF wavefunction at position $\br$ is then
\begin{align}
    \rho(\mathbf r)
    = \sum_{i \in \text{occ}}
      \Big(
        |\psi_i^{\alpha}(\mathbf r)|^2
        + |\psi_i^{\beta}(\mathbf r)|^2
      \Big).
    \label{eq:rho_r_spin_components}
\end{align}
Substituting the orbital expansion from Eq. \ref{eq:psi_alpha}
the electron density at $\br$ can then be written in terms of the diagonal blocks of the density matrix and the real-space atomic orbitals
\begin{align}
    \rho(\bm{r}) = \sum_{\mu\nu} (\bm{P}^{\alpha \alpha}_{\mu\nu} + \bm{P}^{\beta \beta}_{\mu\nu}) \phi_{\mu}(\bm{r})\phi_{\nu} (\bm{r})
\end{align}
Even though the density matrix has $\alpha \beta$ blocks, because the density operator does not depend on spin,
the total GHF electronic density at $\br$ can be expressed solely in terms of the diagonal spin blocks of the density matrix. Furthermore, because $\mathbf{P}$ is Hermitian
($\mathbf{P}^\top = \mathbf{P}^{*}$),
$\rho(\mathbf r)$ depends only on the symmetric real part of $\mathbf{P}$ (and not the antisymmetric imaginary parts of $\mathbf{P}$). 
Electron density plots (along the charge transfer reaction coordinate) for the two active orbitals (i.e., $\bm{P}_{j}= \bm{C}\bm{K}_{j}\bm{C^\dagger}$ where $j \in \{1,2\}$) are drawn in Fig. \ref{fig:cube_soc1}.  These drawing represent the singly and doubly occupied orbitals and are shown for three geometries along the reaction coordinate: the left well minima ( -0.12 \AA), the midpoint ( 0.0 \AA) and the right well (0.12 \AA). A comparative plot is shown in Figure. \ref{fig:cube_soc200} for the strong SOC case ($\eta = 200$).

\section{Discussion: Intruder States with SOC and the Effect of the Constraint}\label{sec:discuss_interuder}

A practical challenge in applying a minimal active-space CASSCF framework to charge transfer is that a variational optimization does not automatically select the desired CT states. Although the target states are those in which the active electron or hole is shared between donor and acceptor fragments, without a constraint, the optimizer may instead converge to  a  local excitation state. We refer to these unwanted solutions as intruder states. If such states enter the active space as the geometry changes, the orbital character can change abruptly, producing non-smooth potential energy surfaces and obscuring the physical meaning of the adiabatic states. 

\begin{figure}[ht]
    \centering
    \includegraphics[width=1.0\textwidth]{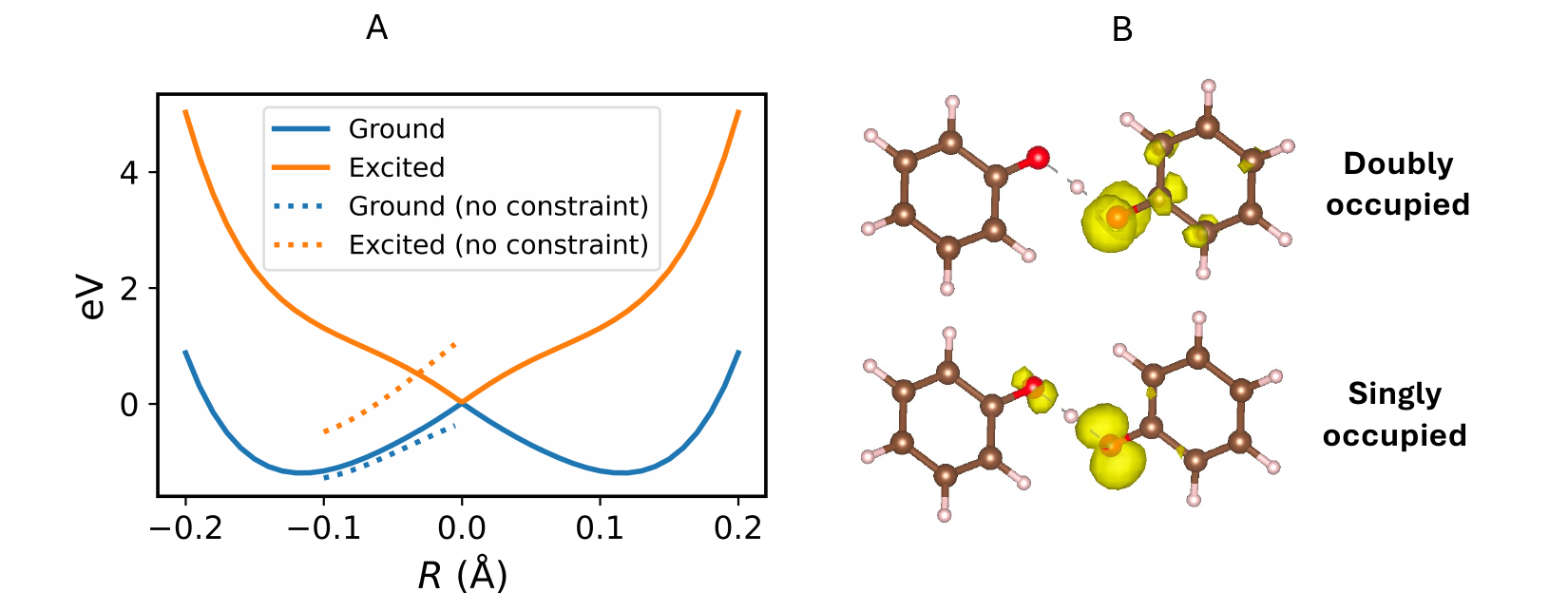}
    \caption{
    Discontinuous PES without constraint. (A) The dotted lines represent the SCF solutions without using the constraint on active space. (B) Electron density in active orbitals. Note that electron/hole is localized on the same fragment in both the orbitals, indicating a local excitation rather than a charge transfer state.}
    \label{fig:intruder} 
\end{figure}

Fig. \ref{fig:intruder}. shows the effect of not including the constraint during the SCF optimization. We end up with discontinuities in the potential energy surfaces which do not represent the CT solutions. 

In the presence of spin--orbit coupling an additional complication arises. Because SOC mixes spin components, the active spinors need not retain a fixed spin character along the reaction coordinate. Near a CT crossing, where several configurations are close in energy, small geometric changes can therefore produce rapid fluctuations in the spin character. We address this spin-gauge ambiguity of doublets and propose a novel diabatization procedure to fix it in Paper 2 \cite{kumar2026chargetransferspinii}.

Numerically, imposing our constraint removes many competing CASSCF solutions that lack the desired CT character. Physically, the constraint enforces a continuous donor--acceptor active space from one well, through the crossing region, to the other well. This continuity will be essential for running nonadiabatic dynamics, and especially so with spin degrees of freedom.

\section{Future Directions and Conclusions}\label{sec:conclusion}
We have presented a method to calculate CT states for molecular systems with an odd number of electrons, keeping track of each state's spin properties.  The end result is a fast and efficient constrained CASSCF(1,2)/CASSCF(3,2) framework for studying charge transfer in the presence of SOC. To that end, we  have investigated a range of SOC strength values and show that energy gap between ground and excited states scales quadratically. While our work here has focused on second row elements (with weak spin-orbit coupling), our approach should be even more important for third row elements, with more SOC and thus a larger entanglement of spin with electron transfer.

Looking forward, one can envision four important future directions. 

\begin{itemize}
    \item 
First, one must wonder: for a four state ($2\times$2) curve crossing, what are the most natural diabatic states that can be constructed so as to preserve the orbital and spin character of a given electronic states. To that end, in a forthcoming article \cite{kumar2026chargetransferspinii}, we will explore different means of constructing diabatic states through adiabatic-to-diabatic transformations where spin cannot be ignored.

\item Second, as we emphasized in the introduction, one would like a means to run dynamics that conserve both the total energy and the total momentum. Now, whereas BO dynamics along a surface $E_{\rm BO}({\bf R}) $(for the most part) ignore electronic momentum,  our research group has shown over the past several years\cite{tao2025basis} that dynamics along a phase space electronic structure surface ($E_{\rm PS}({\bf R}, {\bf P}$) parameterized by both nuclear position and momentum) does conserve electronic momentum. Thus, in the immediate future, it will be quite illuminating to extend the current eDSC/hDSC approach to diagonalize  $H_{\rm PS}({\bf R}, {\bf P})$ rather than $H_{\rm BO}({\bf R}).$  Note that, for the former (as opposed to the latter), the degeneracy of Kramers doublets is lifted because nuclear motion breaks the time reversibility of the electronic Hamiltonian, and so our present algorithm will need to be modified.

\item Third, the method described here is applicable not just for electron transfer between two sites, but also between $N$ sites, so that one can imagine simulating transport across a chain (rather than just transfer between two sites). Indeed, many of the CISS experiments of interest are transport experiments at finite voltage.\cite{gupta2024chirality, bloom2024chiral}

\item Fourth and finally, one would like to explore the extension of the current algorithm to include magnetic fields\cite{bhati2025phase1, bhati2025phase2, rooein2024predicting}, another perturbation that breaks time-reversibility and leads to highly interesting momentum effects. 

\end{itemize}

The bottom line is that, once we begin to explore curve crossing and electron transfer in the presence of  spin degrees of freedom, there is still a great deal of unexplored terrain for theoretical chemistry; the present algorithm is clearly just the tip of the iceberg.

\begin{suppinfo}
Cartesian coordinates for the phenoxy--phenol (ph--ph) system  used to generate the reaction coordinate (PDF)
\end{suppinfo}

\section{Acknowledgments}
This work has been supported by the U.S. Department of Energy, Office of Science, Office of Basic Energy Sciences, under Award no. DE-SC0025393. We have also used resources of the National Energy Research Scientific Computing Center (NERSC), a U.S. Department of Energy Office of Science User Facility operated under Contract no. DE-AC02-05CH11231.

\section{Appendix}

\subsection{Energy gradients w.r.t complex-valued orbital rotation} \label{sec:appendix}

Here, we highlight the energy gradient expression w.r.t the orbital rotation matrix $\bm{A}$. Let us introduce the anti-Hermitian operator
\begin{align}
    \hat{\bm{A}}&= \sum_{pq} A_{pq} a^\dagger_p a_q, \quad \hat{\bm{A}}^\dagger = -\hat{\bm{A}}^\dagger \\
    A_{pq} &= A_{pq}^r + i A_{pq}^i
\end{align}
with $a^\dagger_p (a_q)$ being the creation (annihilation) operator for spin orbital $\chi_p (\chi_q)$. The complex-valued parameters $A_{pq}$ are expressed in terms of its real and imaginary parts. Note that both $A_{pq}^r$ and $A_{pq}^i$ are real parameters and anti-Hermiticity is ensured by requiring them to be anti-symmetric and symmetric, respectively
\begin{subequations}\label{eq:A_symmetry}
\begin{align}
    A_{pq}^r &= - A_{qp}^r \label{eq:A_r}\\
    A_{pq}^i &= A_{qp}^i   \label{eq:A_i}
\end{align}
\end{subequations}
Following ref. \cite{helgakerbook}, the operator ${\bm{\hat{A}}}$ can now be expressed as
\begin{align}
    \hat{\bm{A}} &= \sum_{p>q} A_{pq}^r (a_p^{\dagger}a_q-a_q^{\dagger}a_p) + i \left(  \sum_{p} A_{pp}^i a_p^{\dagger}a_p  + \sum_{p>q} A_{pq}^i (a_p^{\dagger}a_q +a_q^{\dagger}a_p) \right)\\
    &= \sum_{p>q} A_{pq}^r E_{pq}^{-} + i \left(  \sum_{p} A_{pp}^i N_{p}  + \sum_{p>q} A_{pq}^i E_{pq}^{+} \right)
\end{align}

\begin{align}
\left. \frac{\partial E}{\partial A^{r}_{pq}} \right|_{\bm{A}=0}  
&= \left\langle \left[ E^{-}_{pq}, H \right] \right\rangle
= \left( [\bm{F}, \bm{P}]_{qp} - [\bm{F}, \bm{P}]_{pq} \right),
\quad (p > q) \\
\left. \frac{\partial E}{\partial A^{i}_{pq}} \right|_{\bm{A}=0} 
&= i \left\langle \left[ E^{+}_{pq}, H \right] \right\rangle
= i \left( [\bm{F}, \bm{P}]_{qp} + [\bm{F}, \bm{P}]_{pq} \right),
\quad (p > q)
\end{align}
The result for the upper triangle follows directly from the symmetry relations in Eq. \ref{eq:A_symmetry}. 
\begin{equation}
\frac{\partial E}{\partial A^{r}_{qp}}
   = -\frac{\partial E}{\partial A^{r}_{pq}}, \qquad
\frac{\partial E}{\partial A^{i}_{qp}}
   = \;\;\frac{\partial E}{\partial A^{i}_{pq}} .
\end{equation}
The commutator can be simplified in the MO basis which diagonalizes the density matrix. If $\bm{C}$ represents the MO coefficients in the orthonormal AO basis, then $\bm{C}^{\dagger}\bm{C} = \bm{C}\bm{C}^{\dagger} = \bm{I}$. The matrix transformations between the orthonormal AO basis and the MO basis are as follows:
\begin{align}
    \bm{K} &=  \bm{C}^{\dagger}\bm{P}\bm{C}\\
    \bm{F}^{\rm MO} &= \bm{C}^{\dagger}\bm{F}\bm{C}
\end{align}
In the MO basis, the commutator becomes
\begin{align}
    \bm{C}^{\dagger} ([\bm{F},\bm{P}]) \bm{C} = [\bm{F}^{\rm MO},\bm{K}]
\end{align}
where $\bm{K} = \text{diag}(n_1,n_2,..,)$ is a diagonal matrix with $n_j \in \{0,1\}$.
\begin{align}
    ([\bm{F}^{\rm MO},\bm{K}])_{pq} = (n_q-n_p)\bm{F}^{\rm MO}_{pq}
\end{align}

\subsection{N-State Generalization}\label{sec:appendix_nstate}
Although presented above for the case of two charge centers, the extension of a Dynamically-weighted State-Averaged Constrained CASSCF algorithm to the case of $N$ charge centers is straightforward. \cite{qiu2025fast1, qiu2025fast2}. To probe $n^c$ configurations in systems with $n^f$ fragments, one can define the dynamical weighted sum of energies from all configurations, namely,
\begin{align}
    E_{\rm tot} = \sum_{j=1}^{n^c} w_jE_j,\label{eq:E_tot}
\end{align}
where $E_j$ is the energy of the $j$th electronic configuration, $\ket{\Psi_j}$, and $w_j$ is the corresponding weighting factor. In Ref. \cite{qiu2025fast1}, we have proposed that (assuming $\ket{\Psi_1}$ has the lowest energy)
\begin{align}
    w_j &= \frac{a_j}{\sum_j a_j},\\
    a_j &= \frac{1-e^{-\Delta E_j/T}}{\Delta E_j/T},\\
    \Delta E_j &= E_j - E_1.
\end{align}
In terms of the states that we considered, for electron transfer,
\begin{align}
    \ket{\Psi_j} = \ket{1\bar{1},2\bar{2},...,N\bar{N},N+j},\label{eq:edscn_wfc}
\end{align}
which consists the complete active space for one electron in $n^c$ orbitals. For hole transfer,
\begin{align}
    \ket{\Psi_j} = \ket{1\bar{1},2\bar{2},...N-j,\overline{N-j},N-j+1,N-j+2,\overline{N-j+2},...,N\bar{N}},\label{eq:hdscn_wfc}
\end{align}
which consists the complete active space for $2n^c-1$ electron in $n^c$ orbitals. By using the same weighting scheme as proposed in \cite{qiu2024fast}, the differential of Eq. \ref{eq:E_tot} can be written as
\begin{align}
    dE_{\rm tot} = \sum_j w_j'dE_j,
\end{align}
where
\begin{align}
    w_{j>1}' &= \frac{1}{\sum_j a_j}\left(e^{-\Delta E_j/T}-\frac{E_{\rm tot}-E_1}{\Delta E_j}\left(e^{-\Delta E_j/T}-a_j\right)\right),\label{eq:wp_2}\\
    w_1' &= 1-\sum_{j>1}w_j'.\label{eq:wp_1}
\end{align}
As an important ingredient, suppressing local excitations in systems with $n^f$ fragments can be achieved by applying the constraints that 
the vector space spanned by all active orbitals projects equally onto each molecular fragments, i.e.,
\begin{align}
    {\rm Tr}\left(\hat{P}^1\hat{P}^{\rm active}\right) = {\rm Tr}\left(\hat{P}^2\hat{P}^{\rm active}\right) = \cdots = {\rm Tr}\left(\hat{P}^{n^f}\hat{P}^{\rm active}\right).\label{eq:constraints}
\end{align}
where $\hat{P}^{j}$ is the vector space spanned by atomic orbitals from the $j$th molecular fragment and $\hat{P}^{\rm active}$ is the vector space spanned by all active orbitals. This requirement is equivalent to $n^f-1$ constraints:
\begin{align}
    {\rm Tr}\left((\hat{P}^j-\hat{P}^{j+1})\hat{P}^{\rm active}\right) = 0,\ j=1\cdots n^f-1.\label{eq:constraints_v2}
\end{align}
At this point, we can extend the Lagrangian in Eq. \ref{eq:lag_formal} to study CT in systems with multiple charge centers:
\begin{align}
    \mathcal{L} = \sum_{j=1}^{n^c} w_jE_j - \sum_{j=1}^{{n^f}-1}\lambda_j{\rm Tr}\left((\hat{P}^j-\hat{P}^{j+1})\hat{P}^{\rm active}\right).\label{eq:lagrangian_formal}
\end{align}
As an analogue to Eq. \ref{eq:lag_scf}, the gradient of Eq. \ref{eq:lagrangian_formal} w.r.t. the rotation generator $\bm{A}$ can be expressed as
\begin{align}
    \grad{A}\mathcal{L} = [\bm{M}_0^\top,\bm{K}_0+\bar{\bm{K}}_0]+\sum_{j=1}^{n^c}[\bm{M}_j^\top-\bm{M}_{\rm Q}^\top,\bm{K}_j],\label{eq:grad_L_ndscx}
\end{align}
where
\begin{align}
    \bm{M}_{j>0} = w_j'\bm{C}_0^\dagger\bm{F}[\ket{\Psi_j}]\bm{C}_0
\end{align}
is the weighted Fock matrix of state $\ket{\Psi_j}$ in the MO basis, 
\begin{align}
    \bm{M}_0 = \sum_{j=1}^{n^c}\bm{M}_j,
\end{align}
and
\begin{align}
    \bm{M}_{\rm Q} = \sum_{j=1}^{n^f-1}\lambda_j\bm{C}_0^\dagger\bm{Q}_j\bm{C}_0
\end{align}
is the sum of Lagrange multiplier-weighted constraint kernel in the MO basis. Note that $\bm{Q}_j$ represents the difference between vector space spanned by atomic orbitals from the $j$th molecular fragment and the $(j+1)$th molecular fragment in the AO basis. By extending Eq. \ref{eq:lag_scf} to Eq. \ref{eq:grad_L_ndscx}, the same algorithm can be applied to converge the SCF calculations.

\bibliography{cite_all}
\newpage
\begin{tocentry}
  \centering
  \includegraphics[
    width=\linewidth,
    height=1.75in,
    keepaspectratio
  ]{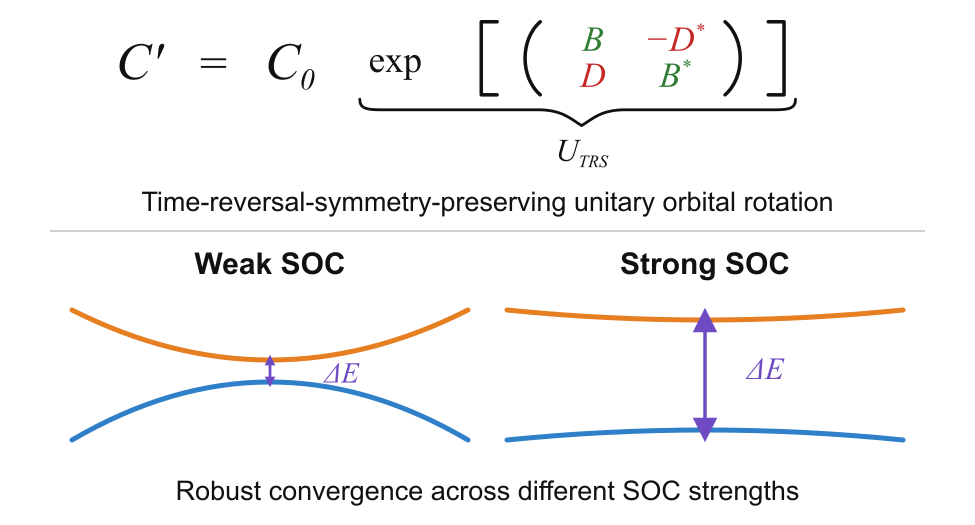}
\end{tocentry}

\end{document}

\renewcommand{\thepage}{S\arabic{page}}
\setcounter{page}{1}

\maketitle

\newpage
\section{Geometry of phenoxy-phenol (ph-ph) system}

% 25\\
% ph-ph\\
 C   -3.339987    1.236679   -0.292676\\
 C   -2.215700    0.395873   -0.461119\\
 C   -2.260068   -0.923462    0.044893\\
 C   -3.395583   -1.374735    0.696165\\
 C   -4.500797   -0.539367    0.860105\\
 C   -4.462623    0.765553    0.361575\\
 O   -1.170747    0.860673   -1.076240\\
 H   -1.404926   -1.560653   -0.084864\\
 H   -3.424139   -2.378976    1.078261\\
 H   -5.377346   -0.897488    1.367665\\
 H   -5.313306    1.410015    0.488105\\
 H   -3.290462    2.235719   -0.683604\\
 C    2.967610   -0.964278   -0.250086\\
 C    1.842996   -0.128779   -0.443279\\
 C    1.884680    1.202903    0.030149\\
 C    3.017817    1.671520    0.673563\\
 C    4.123167    0.841308    0.861788\\
 C    4.087496   -0.476001    0.395845\\
 O    0.800605   -0.609058   -1.049392\\
 H    1.029641    1.836307   -0.118287\\
 H    3.044634    2.685260    1.030351\\
 H    4.998118    1.212844    1.362705\\
 H    4.938888   -1.115655    0.540969\\
 H    2.920157   -1.972439   -0.616477\\
 H   -0.579299    0.419269   -1.076240\\